\RequirePackage{ifpdf}
\documentclass[12pt,a4paper]{JHEP3}
\pdfoutput=1
\usepackage{epsfig}
\usepackage{cite}
\usepackage{amssymb}
\usepackage{amsmath}
\usepackage{xspace}
\def\beq{\begin{equation}}
\def\eeq{\end{equation}}
\def\beqn{\begin{eqnarray}}
\def\eeqn{\end{eqnarray}}
\def\ee{e^+e^-}
\def\as{\alpha_{\rm S}}
\def\bas{\overline{\alpha}_{\rm S}}
\def\Ecm{E_{\rm cm}}
\def\e{{\rm e}}
\def\d{\delta}
\def\k{\kappa}
\def\l{\lambda}

\def\xR{\xi_R}
\def\L{{\cal L}}
\def\N{{\cal N}}
\def\M{{\cal M}}

\def\pd{\partial}
\def\VEV#1{\langle{#1}\rangle}
\def\eq#1{Eq.~(\ref{#1})}
\def\nn{\nonumber\\}

\def\Sherpa{\textsc{SHERPA}\xspace}
\title{\boldmath QCD Jet Rates with the Inclusive Generalized $k_t$ Algorithms}
\author{Erik Gerwick, Steffen Schumann\\
  II. Physikalisches Institut, Universit\"at G\"ottingen, 
  Friedrich-Hund-Platz 1, G\"ottingen 37077, Germany\\
  E-mail: \email{erik.gerwick@phys.uni-goettingen.de}, \email{steffen.schumann@phys.uni-goettingen.de}}
\author{Ben Gripaios, Bryan Webber\\
  Cavendish Laboratory, University of Cambridge, 
   JJ Thomson Avenue, Cambridge CB3 0HE, UK\\
   E-mail: \email{ben.gripaios@cern.ch}, \email{webber@hep.phy.cam.ac.uk}}

\abstract{We derive generating functions, valid to next-to-double
  logarithmic accuracy, for QCD jet rates according to the inclusive
  forms of the $k_ t$, Cambridge/Aachen and anti-$k_ t$ algorithms,
  which are equivalent at this level of accuracy.  We compare the
  analytical results with jet rates and average jet multiplicities
  from the \Sherpa event generator, and study the transition between
  Poisson-like and staircase-like behaviour of jet ratios.
} 

\preprint{Cavendish-HEP-12/18}

\begin{document}
\section{Introduction}
The production of hadronic jets in high-energy collisions is one of
the most striking and universal features of particle physics.  It is
now well understood that jets are associated with hard parton
emission, followed by parton showering and hadronization.  However,
the observed features of jets depend not only on their intrinsic
properties but also on the algorithms used to find them.  These
algorithms have been the object of many years of development and
refinement -- see \cite{Salam:2009jx} for a review.

The jet algorithms used most widely nowadays are of the sequential
recombination type.  Of these, the $k_t$
algorithm~\cite{Catani:1991hj,Catani:1993hr,Ellis:1993tq} was used
extensively at LEP and HERA, and to a limited extent at the Tevatron.
The good theoretical properties of the $k_t$ algorithm mean that
jet rates and multiplicities can be calculated perturbatively,
either in fixed order or in an all-orders logarithmic approximation.
However, the irregular angular shapes of $k_t$-jets make
them less suitable for analyses of hadron collider events, where
multiple parton interactions give rise to underlying activity not
associated with the primary hard process.

A variant of the $k_t$ algorithm that allows better control of the
underlying event is anti-$k_t$~\cite{Cacciari:2008gp}, which has become the preferred jet
finder at the LHC.  It belongs to the general category of inclusive
generalized $k_t$ jet algorithms, defined in~\cite{Cacciari:2011ma}.
Many of the good theoretical features of $k_t$ extend more broadly
to the members of this category.  In particular their leading and
next-to-leading double logarithmic jet rates can be calculated to all
orders, and are in fact the same for all members, including $k_t$ and
anti-$k_t$.

In the present paper we calculate these jet rates, along with the
average jet multiplicities, in a variety of approximations. We do so 
by finding, in each case, the equations that the quark and gluon jet 
generating functions \cite{Konishi:1979cb,Dokshitzer:1991wu,Ellis:1991qj} 
must satisfy. In some cases, an explicit solution in terms of special 
functions can be found, while in others we are forced to resort to
numerical methods (though we can get some analytic understanding of
the properties of the solution, as we describe in detail for one
case in an Appendix). We then compare the results of our theoretical
calculations with those obtained via Monte Carlo simulations using the
\Sherpa event generator \cite{Gleisberg:2008ta,Schumann:2007mg}. Our 
theoretical calculations of jet rates are carried out for the case of
$\ee$ collisions, but we also perform calculations and simulations of 
the average jet multiplicity in hadron-hadron collisions, finding that 
similar behaviour is obtained.

The outline is as follows. We begin by recalling the definitions of
the inclusive jet algorithms and the relevant Sudakov form
factors. In Section~\ref{sec:gen}, we introduce the jet generating
functions and derive the integral equations
that they satisfy at double-leading-logarithmic accuracy (DLA) and
next-to-double-leading-logarithmic accuracy (NDLA). In Sections
\ref{sec:rates} and \ref{sec:mult}, we compute the resulting jet rates
and average jet multiplicities, both at fixed coupling and also at
next-to-leading order in the running coupling. 
In Sections~\ref{sec:mc} and \ref{sec:mcpp}, we compare with results of simulations 
performed using \Sherpa for $\ee$ 
and $pp$ collisions, respectively.  In Section~\ref{sec:sjet} we assess 
the ability for our analytic results to describe sub-jet multiplicities in boosted events.  
In Section~\ref{sec:scale} we consider 
the implications for the scaling patterns of jet multiplicities.  Our 
conclusions are summarized in Section~\ref{sec:conc}.  Details of the
derivation and properties of the partial differential equation (PDE)
for the average jet multiplicity are relegated to appendices.

\section{\boldmath The inclusive generalized $k_t$ jet algorithms\label{sec:inc}}
We consider first the case of multijet production in $\ee$
annihilation, for which the inclusive algorithms are defined as described
in the FastJet user manual~\cite{Cacciari:2011ma}, Sect.~4.5. 
The distance measures are
\beqn\label{eq:genktee}
d_{ij}
&=&\min\{E_i^{2p},E_j^{2p}\}\frac{(1-\cos\theta_{ij})}{(1-\cos R)}\nn
&\equiv&\min\{E_i^{2p},E_j^{2p}\}\xi_{ij}/\xR\;,\nn
d_{iB} &=& E_i^{2p}\;,
\eeqn
with $p=1,0,-1$ for the $k_ t$~\cite{Catani:1991hj},
Cambridge/Aachen~\cite{CamOrig,CamWobisch} and
anti-$k_ t$~\cite{Cacciari:2008gp} algorithms, respectively.
At any stage of clustering, if a $d_{ij}$ is the smallest measure we
combine objects $i$ and $j$.  If $d_{iB}$ is the smallest we call $i$ a jet
candidate and remove it from the clustering list.  We then call jet
candidates with energy $E_i>E_R$ resolved jets.  Thus the jet rates,
at a given value of the $\ee$ centre-of-mass energy $\Ecm$,
are functions of the radius parameter $R$ and the minimum jet energy
$E_R$.  This is in contrast to the {\em exclusive} $k_t$ (Durham)
algorithm, where one effectively sets $\xR=\frac 12$ and continues
clustering objects  until all $d_{ij}$ are above some fixed value
$d_{\rm cut}=y_{\rm cut}\Ecm^2$, so that the jet rates are functions
of the single parameter $y_{\rm cut}$.

In hadron-hadron collisions, the c.m.\ frame of the  parton-parton
hard scattering process is not known and therefore one has to adopt a
longitudinally invariant form of the
algorithms~\cite{Catani:1993hr,Ellis:1993tq}.
To that end, Eqs.~(\ref{eq:genktee}) are replaced by
\beqn\label{eq:genktpp}
d_{ij}
&=&\min\{p_{ti}^{2p},p_{tj}^{2p}\}\frac{\Delta R_{ij}^2}{R^2}\;,\nn
\Delta R_{ij}^2 &\equiv& (y_i-y_j)^2+(\phi_i-\phi_j)^2\;,\nn
d_{iB} &=& p_{ti}^{2p}\;,
\eeqn
where $p_{ti}$, $y_i$ and $\phi_i$ are the transverse momentum,
rapidity and azimuth of object $i$, respectively, and we define
jet candidates with $p_{ti}>E_R$ as resolved jets.

As far as leading logarithms are concerned, the jet rates defined by
(\ref{eq:genktee}) and (\ref{eq:genktpp}) will be the same, and
therefore in the following Sections we refer mainly to
$\ee$ annihilation.  By ``leading logarithms'' here we always mean leading
double and next-to-double logarithms, $\as^n\log^{2n}$ and
$\as^n\log^{2n-1}$, where the logarithms are those of $1/R$ and/or
$Q/E_R$, $Q$ being the hard process scale.   By taking $E_R$
sufficiently large in hadron-hadron collisions, we avoid such 
leading contributions from initial-state showering and the underlying 
event, so these terms are determined by the timelike showering of 
final-state partons.

\section{Sudakov factors\label{sec:sud}}
The evolution scale for coherent parton showering is
$\xi\equiv 1-\cos\theta$ with $\theta$ the emission angle.
To be resolved, an emission must have $\xi>\xR$
and $E>E_R$. The probability for a single resolvable gluon emission from
a quark of energy $E$ at scale $\xi$ is thus
\beq\label{eq:nobran}
{\cal P}_q(E,\xi)=\int_{\xR}^{\xi} \frac{d\xi'}{\xi'}
\int_{E_R/E}^1 dz\frac{\as(k_t^2)}{2\pi}P_{gq}(z)\;,
\eeq
where the running coupling is evaluated at the transverse momentum
scale of the emission, $k_t^2=z^2E^2\xi'$, 
\beq
\frac{\as(k_t^2)}{\pi} = \frac 1{b_0\ln(z^2E^2\xi'/\Lambda^2)}
\eeq
with $b_0=(11C_A-2n_f)/12$.  Defining $\bas = \as(E^2\xi)/\pi$, i.e.\
in terms of the coupling at the hard scale,
we have to  next-to-double-log accuracy (NDLA)
\beq\label{eq:askt}
\frac{\as(k_t^2)}{\pi}  = \bas -b_0\bas^2\left[2\ln z+\ln\left(\frac{\xi'}{\xi}\right)\right]
\eeq
and
\beqn\label{eq:PqExi}
{\cal P}_q(E,\xi)&=&
C_F\bas\ln\left(\frac\xi\xR\right)\left[\ln\left(\frac
    E{E_R}\right)-\frac 34\right] +\nn
&&\frac 12 C_F b_0\bas^2\ln\left(\frac\xi\xR\right)\ln\left(\frac E{E_R}\right)
\left[2\ln\left(\frac
    E{E_R}\right)+\ln\left(\frac\xi\xR\right)\right]\;.
\eeqn
Then the probability for no resolvable emissions (the quark Sudakov factor) is
\beq
\Delta_q(E,\xi) = \exp\left[-{\cal P}_q(E,\xi)\right]\;.
\eeq
Similarly for a gluon, the probability of a single resolvable gluon,
quark or antiquark emission is
\beq
{\cal P}_g(E,\xi)=\int_{\xR}^{\xi} \frac{d\xi'}{\xi'}
\int_{E_R/E}^1 dz\frac{\as(k_t^2)}{2\pi}\left[P_{gg}(z)+P_{qg}(z)\right]\;,
\eeq
which gives to NDLA
\beqn\label{eq:PgExi}
{\cal P}_g(E,\xi)&=&
\bas\ln\left(\frac\xi\xR\right)\left[C_A\ln\left(\frac E{E_R}\right)-b_0\right] +\nn
&&\frac 12 C_A b_0\bas^2\ln\left(\frac\xi\xR\right)\ln\left(\frac E{E_R}\right)
\left[2\ln\left(\frac
    E{E_R}\right)+\ln\left(\frac\xi\xR\right)\right]\;,
\eeqn
and the gluon Sudakov factor is
\beq
\Delta_g(E,\xi) = \exp\left[-{\cal P}_g(E,\xi)\right]\;.
\eeq
Note that all this is independent of the value of $p$, so that all the
 inclusive generalized $k_t$ algorithms are equivalent at this level of precision.

\section{Generating functions\label{sec:gen}}
By definition the generating function for resolved jets from a quark
($i=q$) or gluon ($i=g$) of energy
$E$ at scale $\xi$ is~\cite{Konishi:1979cb,Dokshitzer:1991wu,Ellis:1991qj}
\beq
\Phi_i(u,E,\xi) = \sum_{n=0}^\infty u^n\,R_n^i(E,\xi)\;,
\eeq
where $R_n^i$ is the corresponding $n$-jet rate, i.e.\ the probability of finding
$n$ resolved jets.  Thus the jet rates can be recovered from the generating
function by successive differentiation at $u=0$:
\beq
R_n^i(E,\xi) = \frac 1{n!}\frac{\pd^n}{\pd
  u^n}\Phi_i(u,E,\xi)|_{u=0}\;.
\eeq
On the other hand the average multiplicity of resolved jets is
obtained by differentiating at $u=1$. Writing the average jet
multiplicity from a quark or gluon of energy
$E$ at scale $\xi$ as $\N_i(E,\xi)$, we have
\beq
\N_i(E,\xi) =  \sum_{n=0}^\infty n\,R_n^i(E,\xi)
=\frac{\pd}{\pd u}\Phi_i(u,E,\xi)|_{u=1}\;.\label{eq:avmult}
\eeq
The generating functions $\Phi_{q,g}$ must thus satify the
boundary condition
\beq
\Phi_i(u,E,\xR) = 1+(u-1)\Theta(E-E_R)\;.
\eeq

The generating function for $\ee$ annihilation at c.m.\ energy $\Ecm$
is that for two quarks of energy $\Ecm/2$, each filling one hemisphere:
\beq\label{eq:Phiee}
\Phi_{ee}= [\Phi_q(u,\Ecm/2,1)]^2\;.
\eeq

\subsection{Next to double leading logarithms}
For $\xi>\xR$ and $E>E_R$, we have to NDLA
\beqn
\Phi_q(u,E,\xi) &=& u\Delta_q(E,\xi) +\int_{\xR}^{\xi} \frac{d\xi'}{\xi'}\frac{\Delta_q(E,\xi)}{\Delta_q(E,\xi')}
\int_{E_R/E}^1 dz\frac{\as(k_t^2)}{2\pi}P_{gq}(z)\Phi_q(u,E,\xi')\Phi_g(u,zE,\xi')\;,\nn
\Phi_g(u,E,\xi) &=& u\Delta_g(E,\xi) +\int_{\xR}^{\xi} \frac{d\xi'}{\xi'}\frac{\Delta_g(E,\xi)}{\Delta_g(E,\xi')}\int_{E_R/E}^1
dz\frac{\as(k_t^2)}{2\pi}\bigl\{P_{gg}(z)\Phi_g(u,E,\xi')\Phi_g(u,zE,\xi')\nn
&&+ P_{qg}(z)[\Phi_q(u,E,\xi')]^2\bigr\}\;.
\eeqn

Using the expressions given earlier to eliminate the Sudakov factors, we may write these in
the equivalent forms
\beqn
\Phi_q(u,E,\xi) &=& u +\int_{\xR}^{\xi} \frac{d\xi'}{\xi'}
\int_{E_R/E}^1 dz\frac{\as(k_t^2)}{2\pi}P_{gq}(z)\Phi_q(u,E,\xi')
\left[\Phi_g(u,zE,\xi')-1\right]\;,\nn
\Phi_g(u,E,\xi) &=& u +\int_{\xR}^{\xi} \frac{d\xi'}{\xi'}
\int_{E_R/E}^1dz\frac{\as(k_t^2)}{2\pi}\bigl\{P_{gg}(z)\Phi_g(u,E,\xi')
\left[\Phi_g(u,zE,\xi')-1\right]\nn
&&+ P_{qg}(z)\left[\{\Phi_q(u,E,\xi')\}^2-\Phi_g(u,E,\xi')\right]\bigr\}\;.
\eeqn

The solution for the quark generating function is then easily seen to be
\beq
\Phi_q(u,E,\xi) = u\exp\left\{\int_{\xR}^{\xi} \frac{d\xi'}{\xi'}
\int_{E_R/E}^1 dz\frac{\as(k_t^2)}{2\pi}P_{gq}(z)
\left[\Phi_g(u,zE,\xi')-1\right]\right\}\;.
\eeq
We can solve for the gluon generating function by iteration,
and then substitute in this equation to get the complete solution.

\subsection{Leading double logarithms}
In the DLA we keep only the singular parts of $P_{gq}$ and $P_{gg}$
and can drop $P_{qg}$.  For brevity we define the logarithms as
\beq\label{eq:kldef}
\k =\ln(E/E_R)\;,\;\;\;
\l =\ln(\xi/\xR)\;.
\eeq
Then
\beq\label{eq:SudDLA}
\Delta_{q,g}(\k,\l) = \e^{-a_{q,g}\k\l}
\eeq
where $a_{q,g} = C_{F,A}\bas$, and
\beqn\label{eq:PhiDLA}
\Phi_q(u,\k,\l) &=& u\,\e^{-a_q\k\l}\exp\left\{a_q\int_0^\k d\k' \int_0^\l d\l'
\,\Phi_g(u,\k',\l')\right\}\;,\\
\Phi_g(u,\k,\l) &=&  \e^{-a_g\k\l}\left\{u+a_g\int_0^\k d\k' \int_0^\l d\l'
\,\e^{a_g\k\l'}\Phi_g(u,\k,\l')\Phi_g(u,\k',\l')\right\}\;.\nonumber
\eeqn
We can simplify the equation for $\Phi_g$ by noting that
\beq
\frac{\pd}{\pd\l}\left(\e^{a_g\k\l}\Phi_g\right)
=a_g\e^{a_g\k\l}\Phi_g\int_0^\k d\k'\,\Phi_g(u,\k',\l)
\eeq
so that
\beq
\ln\left(\e^{a_g\k\l}\Phi_g\right)
=\int_0^\k d\k'\int_0^\l d\l'\,\Phi_g(u,\k',\l') + C(u,\k)
\eeq
where the boundary condition $\Phi_g(u,\k,0)=u$ for all $\k\ge 0$
implies $C(u,\k)=\ln u$.  Thus
\beq\label{eq:PhigDLA}
\Phi_g(u,\k,\l) =  u\,\e^{-a_g\k\l}\exp\left\{a_g\int_0^\k d\k' \int_0^\l d\l'
\,\Phi_g(u,\k',\l')\right\}\;.
\eeq
Comparing with (\ref{eq:PhiDLA}) we see that in the DLA
\beq\label{eq:Phiqg}
\Phi_q = u\left(\Phi_g/u\right)^{C_F/C_A}\;.
\eeq

\section{Jet rates\label{sec:rates}}
From the quark and gluon generating functions in the previous 
section it is relatively straightforward to construct the jet rates
to next-to-double log accuracy (NDLA). 
In terms of the logarithmic variables (\ref{eq:kldef})
the quark and gluon integrated splitting 
kernels (\ref{eq:PqExi}) and (\ref{eq:PgExi}) are simply
\beq
{\cal P}_q(E,\xi)=
\bas \l C_F\left[\kappa -\frac{3}{4}\right] + 
\frac 12 C_F b_0\bas^2 \l \k
\left[2 \k + \l \right]\;,
\eeq
\beq
{\cal P}_g(E,\xi)=
\bas \l \left[C_A \kappa -b_0\right] + 
\frac 12 C_A b_0\bas^2 \l \k
\left[2 \k + \l \right]\;,
\eeq
where
\beq\label{eq:bas}
\bas \equiv \bas(\k,\l) = \frac 1{b_0(2\k+\l+\mu)}
\eeq
with $\mu=\ln(E_R^2\xi_R/\Lambda^2)$.

The quark generating function is
\beqn\label{eq:Phiq}
\Phi_q(u,E,\l) 
    & = & u\exp\left\{C_F \int_{0}^{\l} {d\l'}
\int_{0}^{\k} 
d\k' \,\left( \bas(\k',\l')  -  \frac{3}{4} \bas 
\e^{\k'-\k} \right)\left[\Phi_g(u,\k',\l')-1\right]\right\}\nn
\eeqn
where, as in (\ref{eq:askt}), to NDLA we have
$\bas(\k',\l') =\bas -b_0\bas^2\left(2\k'-2\k+\l'-\l\right)$.
In the fixed-order expansion it is not necessary to 
include the running coupling for the finite terms in the 
splitting functions, as doing so would affect results beyond the NDLA.  In 
these finite term we can set terms proportional 
to $e^{-\k}$ to 0 after the integrations, which is equivalent to 
allowing the original $z$ integration to range over $(0,1)$, since it is not 
singular in energy.  However, we have to bear in mind that
the relevant integrals vanish when $\k=0$.

Defining 
\beqn
\Gamma_g(\k',\l',\k) &=&  C_A \left[ \bas (\k',\l') 
- \frac{11}{12} \bas  \e^{\k'-\k}  \right] \;,\\
\Gamma_q(\k',\l',\k) &=&  C_F \left[ \bas(\k',\l') 
- \frac{3}{4} \bas  \e^{\k'-\k}  \right]\;,
\eeqn
we solve the gluon generating function by iteration to third order in $u$.
\begin{alignat}{5} 
\Phi^{(1)}_{g} (u,\k, \l) &= u \Delta_g(\k, \l) \\
\Phi^{(2)}_{g} (u,\k, \l) &= u \Delta_g(\k, \l) \left(1 + u  
\int_0^{\l} d \l' \int_0^{\k} d\k'  
\;\bigg\{ \Gamma_g(\k',\l',\k) \,
\Delta_g(\k', \l') 
\right.  \notag \\
& +  \left. \left. \bas \frac{n_f}{6} \frac{\Delta^2_q(\k, \l')}{\Delta_g(\k, \l')} \e^{\k'-\k} \right\} 
\right)    \\  
\Phi^{(3)}_{g} (u,\k, \l) 
&= u \Delta_g(\k, \l) \left(1 + u  
\int_0^{\l} d \l' \int_0^{\k} d\k' \bigg\{ \Gamma_g(\k',\l',\k)  
\Delta_g(\k', \l') 
\right.   \notag \\
 \times  & \left( 1 +  u \,  \left[ \int_0^{\k} + \int_0^{\k'} \right] d \k'' \int_{0}^{\l'}
d\l''\; \bigg\{ \Gamma_g(\k'',\l'',\bar{k})   \Delta_g(\k'', \l'')  \right. 
 \notag \\
&  + \, \left. \left. \bas \frac{n_f}{6} \frac{\Delta^2_q(\k', \l'')}{\Delta_g(\k', \l'')} \e^{\k''-\bar{\k}} \right\} \right)
\notag \\
&+ \:  \bas \frac{n_f}{6} \frac{\Delta^2_q(\k, \l')}{\Delta_g(\k, \l')} \e^{\k'-\k} 
\bigg( 1 + 2u \,\int_{0}^{\l'} d \l'' \int_{0}^{\k} d \k''  
\notag \\ 
& \times  \Gamma_q(\k'',\l'',\k)  \Delta_g(\k'', \l'') 
\bigg)
\bigg\} 
\bigg)\,.
\end{alignat}
In the result for $\Phi^{(3)}_{g} $ we have defined $\bar{\k} = \k  (\k')$ 
when the $\k''$ integral ranges from $0$ to $\k$ $(\k')$.  Defining
\beq
\Psi_g^{(n)} = 2 \int_{0}^{\k} d\k' \int_{0}^{\l} d\l'  \; \Gamma_q(\k',\l',\k) \, \Phi^{(n)}_{g} (u,\k', \l')\,,
\eeq
the generating function for resummed rates up to 5 jets at NDLA is then
\beq
\Phi_{ee}^{(5)} = u^2 \Delta_{q}^{2}(\k,\l) \left\{ 
1 + \Psi_g^{(3)} + \frac{1}{2} \left(\Psi_g^{(2)}\right)^2 + \frac{1}{3!} \left(\Psi_g^{(1)}\right)^3 \right\}\,.
\label{resummed_func5}
\eeq
Substituting in (\ref{eq:Phiq}) and using (\ref{eq:Phiee}),
we find the rates for $\ee$ annihilation in the form
\beq
R_n^{ee} = \delta_{2,n}+\sum_{j\ge n-2}\bas^j\left(R_{n,2j}+R_{n,2j-1}\right)
\eeq
for $n\le 5$ and $j\le 3$, where  $n$ is the number of resolved jets and 
 $R_{n,i}$ has $i$ powers of the logarithms (either $\k$ or $\l$): 
\beqn\label{eq:R2}
R_{22} &=& - 2 C_F \k \l \nn
R_{21} &=& \;\; \frac{3}{2} C_F \l \nn
R_{24} &=& \;\; 2 C_F^2 \k^2 \l^2 \nn
R_{23} &=& \left[-b_0 ( 2 \k + \l)  - 3 C_F\l\right]C_F\k\l \nn
R_{26} &=& -\frac{4}{3} C_F^3 \k^3 \l^3 \nn
R_{25} &=& \left[2 b_0 ( 2\k + \l)+3 C_F\l\right] C_F^2\k^2 \l^2 \\
&& \nn
R_{32} &=& \;\; 2 C_F \k \l \nn
R_{31} &=& -\frac{3}{2} C_F \l \nn
R_{34} &=& \left[- 4 C_F -  \frac{1}{2} C_A \right]C_F\k^2 \l^2 \nn
R_{33} &=&  \left[b_0 ( 2 \k + \l) + \left(\frac 53 C_A +6 C_F
    - \frac 16 n_f\right)\l\right]C_F\k \l \nn
R_{36} &=&  \left[ 4 C_F^2  + C_A C_F + \frac{1}{9} C_A^2 \right] C_F\k^3 \l^3 \nn
R_{35} &=& \biggl[\left(
-\frac{1}{2}C_A - 4 C_F \right)b_0(2\k + \l) +\nn &&
\biggl(-\frac{5}{9} C_A^2-\frac{49}{12} C_A C_F - 9 C_F^2+\frac{1}{18} 
C_A n_f  + \frac{1}{3} C_F n_f\biggr)\l\biggr]C_F\k^2 \l^2 
\eeqn
\beqn\label{eq:R4}
R_{44} &=& \left[ 2 C_F + \frac{1}{2} C_A\right]C_F \k^2 \l^2  \nn
R_{43} &=& \left[- \frac 53 C_A -3C_F +\frac 16 n_f\right]C_F\k \l^2  \nn
R_{46} &=& \left[- 4 C_F^2 - 2 C_A C_F -\frac{5}{18} C_A^2\right]C_F \k^3 \l^3  \nn
R_{45} &=& 
\biggl[\left(\frac{1}{2} C_A +2C_F\right) b_0(2 \k + \l) 
+ \nn &&
\biggl(\frac{37}{24} C_A^2 +  \frac{49}{6} C_A C_F
+ 9 C_F^2 -\frac{1}{9} C_A n_f - \frac{13}{18} C_F n_f\biggr)\l\biggr]C_F\k^2 \l^2 
\eeqn
\beqn\label{eq:R5}
R_{56} &=& \left[  \frac{4}{3} C_F^2 + C_A C_F  + \frac{1}{6} C_A^2\right] C_F\k^3 \l^3
\nn
R_{55} &=& \left[
-\frac{71}{72} C_A^2 - \frac{49}{12} C_A C_F- 3 C_F^2  + 
 \frac{1}{18} C_A n_f + \frac{7}{18} C_F n_f\right] C_F \k^2 \l^3
\eeqn
One check on these results is that the DLA coefficients agree with the previous computation in 
\cite{Webber:2010vz} for $R_{44}$ and $R_{56}$.  A second is that the 2-jet 
inclusive fraction obtained by summing the rates is 1.

\section{Average jet multiplicity\label{sec:mult}}
The average jet multiplicity, i.e.\ the mean number of resolved jets, as 
a function of the hard process scale $Q$, the angular resolution $R$ and 
the minimum jet energy $E_R$, provides a useful overall measure of jet 
activity and substructure.  As indicated by \eq{eq:avmult}, this 
quantity is obtained simply from the first derivative of the relevant 
generating function.
\subsection{Next to double leading logarithms}
 Writing the average jet multiplicity in $\ee$ annihilation as
 $\N_{ee}$, from the fixed-order jet rates (\ref{eq:R2})-
 (\ref{eq:R5}) we obtain to $\mathcal{O}(\as^3)$
\beqn\label{av_fo}
\N_{ee}  & = & 2 + \bas \left(2 \k \l 
-\frac{3}{2}\l\right)C_F
\nn &+& \bas^2 \left(\frac{1}{2} C_A\k \l +  2 b_0 (2\k  +  \l) 
 - \frac{5}{3} C_A \l +  \frac{1}{6} n_f \l \right) C_F\k \l
\nn &+& \bas^3 \left(\frac{1}{18} C_A^2\k \l
+  \frac{1}{2}b_0 C_A(2\k  + \l) 
+  \frac{1}{18} n_f C_F \l
- \frac{31}{72} C_A^2  \l \right)C_F\k^2 \l^2\;.
\eeqn
Here the terms in $b_0$ originate from including the running coupling.  
We see that these terms enhance the average jet multiplicity with 
respect to a fixed-coupling calculation.  

To perform an all-orders resummation to NDLA,
we repeat the analysis of \cite{Catani:1991pm}, for the inclusive algorithms
 instead of the exclusive $k_t$ algorithm.  In terms of the generating
 functions, we have
\beq
\N_{ee}(E,\xi) =\left.\frac{\pd\Phi_{ee}}{\pd u}\right|_{u=1}
=2\left.\frac{\pd\Phi_q}{\pd u}\right|_{u=1} = 2\N_q
\eeq
where $\N_{q,g}$, the average quark and gluon jet multiplicities, satisfy the equations
\beqn\label{eq:Nq}
\N_q(E,\xi) &=&  1+\int_{\xR}^{\xi} \frac{d\xi'}{\xi'}
\int_{E_R/E}^1 dz\frac{\as(k_t^2)}{2\pi}P_{gq}(z)\N_g(zE,\xi')\\
\label{eq:Ng}
\N_g(E,\xi) &=& 1 +\int_{\xR}^{\xi} \frac{d\xi'}{\xi'}
\int_{E_R/E}^1dz\frac{\as(k_t^2)}{2\pi}\bigl\{P_{gg}(z)\N_g(zE,\xi')\nn
&&+ P_{qg}(z)\left[2\N_q(E,\xi')-\N_g(E,\xi')\right]\bigr\}\;.
\eeqn

In Appendix~\ref{sec:PDEderiv} we show that (\ref{eq:Nq}) is
equivalent to the following PDE
in terms of the logarithmic variables (\ref{eq:kldef}):
\beq\label{eq:dNqdkdl}
\frac{\pd^2\N_q}{\pd\k\pd\l} 
=C_F\bas\left(\N_g-\frac 34 \frac{\pd\N_g}{\pd\k}\right)\;,
\eeq
with boundary conditions $\N_q(\k,0)=\N_q(0,\l)=1$.

Similarly we find from (\ref{eq:Ng})
\beq
\frac{\pd^2\N_g}{\pd\k\pd\l} 
=\bas\left[C_A\N_g-\left(\frac{11}{12}C_A+\frac{n_f}6\right) \frac{\pd\N_g}{\pd\k}\right]
+\frac{n_f}3\bas\frac{\pd\N_q}{\pd\k}\;,
\eeq
where to the required accuracy we may set in the last term
\beq
\frac{\pd\N_q}{\pd\k}=\frac{C_F}{C_A}\frac{\pd\N_g}{\pd\k}\;,
\eeq
so that finally
\beq\label{eq:dNgdkdl}
\frac{\pd^2\N_g}{\pd\k\pd\l} 
=\bas\left[C_A\N_g-\left(\frac{11}{12}C_A+\frac{n_f}6-\frac{n_fC_F}{3C_A}\right)
\frac{\pd\N_g}{\pd\k}\right]\;,
\eeq
with boundary conditions $\N_g(\k,0)=\N_g(0,\l)=1$.

Note that the $n_f$ dependence in (\ref{eq:dNgdkdl}) is very
weak and vanishes in the large-$N$ limit:
\beq
\frac{n_f}6-\frac{n_fC_F}{3C_A} = \frac{n_f}{6N^2}=\frac{n_f}{54}\;.
\eeq
This is because at large $N$ a $q\bar q$ pair from gluon splitting
radiates like a gluon.

\subsection{Leading double logarithms}
Dropping the non-singular parts of the splitting functions, we have
from (\ref{eq:Phiqg})
\beq
\N_{ee} =2\N_q = 2+2\frac{C_F}{C_A}\left(\N_g-1\right)
\eeq
where
\beq\label{eq:NgDLA}
\frac{\pd^2\N_g}{\pd\k\pd\l}=C_A\bas\N_g\;.
\eeq

In the leading double log approximation, $\as$ is treated as a
constant. Then the solution
to (\ref{eq:NgDLA}) is a modified Bessel function:
\beq\label{eq:Avn_I0}
\N_g(\k,\l) = \sum_{n=0}^\infty
\frac{(C_A\bas\k\l)^n}{(n!)^2} = I_0\left(2\sqrt{C_A\bas\k\l}\right)\;.
\eeq

The asymptotic behaviour for large argument,
\beq
I_0(y) \sim \frac {\e^y}{\sqrt{2\pi y}}\;,
\eeq
implies that for high energy and small cone size
\beqn\label{eq:Avn_asy}
\N_{ee} \sim 2\left(1-\frac{C_F}{C_A}\right) &+& \frac{C_F}{\sqrt{\pi}C_A}
\left[C_A\bas\ln\left(\frac \Ecm{2E_R}\right)
\ln\left(\frac 1{\xR}\right)\right]^{-\frac 14}\times\nn
&&\exp\left[2\sqrt{C_A\bas\ln\left(\frac \Ecm{2E_R}\right)
\ln\left(\frac 1{\xR}\right)}\right]\,.
\eeqn

\subsection{Running coupling}
Taking into account the running of $\bas$ to
next-to-leading order, we have
\beq
\frac{\pd^2}{\pd\k\pd\l}I_0\left(2\sqrt{C_A\bas\k\l}\right)=
[1-b_0(2\k+\l)\bas+{\cal O}(\as^2)]I_0\left(2\sqrt{C_A\bas\k\l}\right)\;.
\eeq
Thus if we drop terms of relative order $\as^2$ the solution to (\ref{eq:NgDLA}) is
\beq
N_g = [1+b_0(2\k+\l)\bas]\,I_0\left(2\sqrt{C_A\bas\k\l}\right)\;,
\eeq
which agrees with the $b_0$-dependent terms in
(\ref{av_fo}).  However, for large $\k$ and/or $\l$,
$b_0(2\k+\l)\bas\sim 1$ and therefore we need to take
into account the running of $\as$ to all orders. 

\subsection{Numerical solution: DLA}
 Treating the running of $\as$ to all orders, as in
 \eq{eq:bas}, but still neglecting the
 finite parts of the splitting functions,
 we have in place of (\ref{eq:NgDLA})
\beq\label{eq:NgPDE}
\frac{\pd^2\N_g}{\pd\k\,\pd\l}= c_g\frac{\N_g}{(2\k+\l+\mu)}\;,
\eeq
with  $c_g = C_A/b_0$.
This PDE is not straightforward to solve explicitly.  Its properties
are discussed in Appendix~\ref{sec:PDEprops}.  It may however
be solved numerically by discretization. Writing
\beqn
\N_q(\k=ma,\l=nb)&=&f_{m,n}\nn
\N_g(\k=ma,\l=nb)&=&g_{m,n}\;,
\eeqn
we have   
\beq\label{eq:Nglhs}
\frac{\pd^2\N_g}{\pd\k\,\pd\l}\approx
\frac 1{ab}\left[g_{m+1,n+1}-g_{m+1,n}-g_{m,n+1}+g_{m,n}\right]
\eeq
and
\beqn\label{eq:Ngrhs}
\frac{c_g\N_g}{(2\k+\l+\mu)}&\approx&
\frac{c_g}4\biggl[\frac{g_{m+1,n+1}}{2(m+1)a+(n+1)b+\mu}
+\frac{g_{m+1,n}}{2(m+1)a+nb+\mu}\nn
&&+\frac{g_{m,n+1}}{2ma+(n+1)b+\mu}
+\frac{g_{m,n}}{2ma+nb+\mu}\biggr]\;.
\eeqn
Equating these expressions, one can solve iteratively for
$g_{m+1,n+1}$ starting from the boundary values $g_{0,n}=g_{m,0}=1$.

\subsection{Numerical solution: NDLA}
To include the finite parts of the splitting functions, we may write
(\ref{eq:dNqdkdl}) and (\ref{eq:dNgdkdl}) with equivalent precision as
\beq\label{eq:Ngcd}
\frac{\pd^2\N_{q,g}}{\pd\k\pd\l}
=c_{q,g}\left(1-d_{q,g}\frac{\pd}{\pd\k}\right)\frac{\N_g}{2\k+\l+\mu}\;,
\eeq
where $c_{q,g}=C_{F,A}/b_0$ and 
\beq
d_q=\frac 34\;,\;\;\;
d_g=\frac{11}{12}+\frac{n_f}{6N^3}\;.
\eeq

The PDEs (\ref{eq:Ngcd}) can be solved numerically by a simple
extension of the method
outlined above. For the discretized $\k$-derivative, we use
\beq
\frac{\pd\N_g}{\pd\k}\approx
\frac 1{2a}\left[g_{m+1,n+1}+g_{m+1,n}-g_{m,n+1}-g_{m,n}\right]\;.
\eeq
We can then write the right-hand side of (\ref{eq:Ngcd}) as
\beqn\label{eq:NgrhsNDLA}
&&\frac{c_g}4\biggl[\frac{(1-\d_g)g_{m+1,n+1}}{2(m+1)a+(n+1)b+\mu}
+\frac{(1-\d_g)g_{m+1,n}}{2(m+1)a+nb+\mu}\nn
&&\qquad +\frac{(1+\d_g)g_{m,n+1}}{2ma+(n+1)b+\mu}
+\frac{(1+\d_g)g_{m,n}}{2ma+nb+\mu}\biggr]\;,
\eeqn
where
\beq
\d_g=\frac 2a d_g = \frac 2a\left(\frac{11}{12}+\frac{n_f}{6N^3}\right)\;,
\eeq
and equate this to (\ref{eq:Nglhs}).

Similarly, to obtain the quark jet multiplicity we write   
\beq\label{eq:Nqlhs}
\frac{\pd^2\N_q}{\pd\k\,\pd\l}\approx
\frac 1{ab}\left[f_{m+1,n+1}-f_{m+1,n}-f_{m,n+1}+f_{m,n}\right]\;,
\eeq
equate this to (\ref{eq:NgrhsNDLA}) with $c_g,\d_g$ replaced by
\beq
c_q= \frac{C_F}{b_0}\,,\;\;\; \d_q=\frac 3{2a}
\eeq
to obtain the discrete equivalent of (\ref{eq:dNqdkdl}),
and solve iteratively for $f_{m+1,n+1}$ starting from the boundary
values $f_{0,n}=f_{m,0}=1$.

\section{\boldmath Monte Carlo results: $e^+ e^-$\label{sec:mc}}
\FIGURE{
  \centering\centerline{
  \includegraphics[scale=0.8]{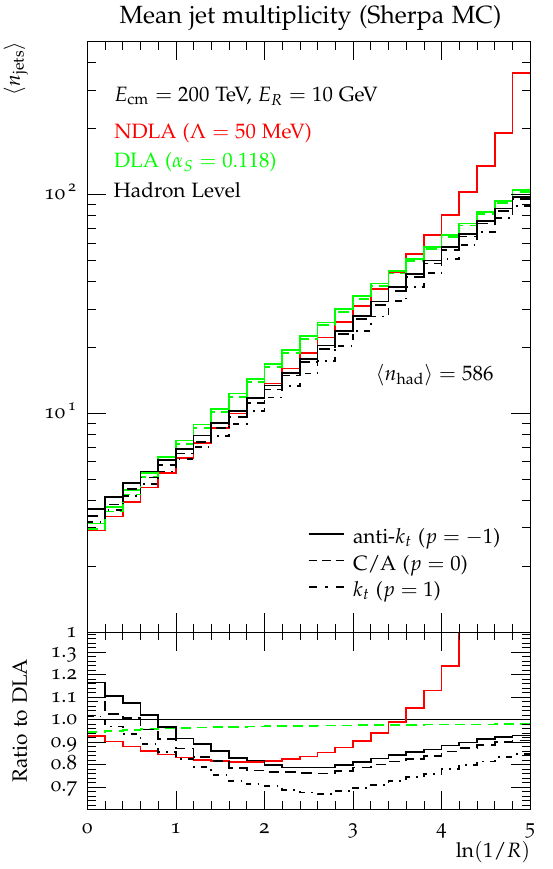}
  \includegraphics[scale=0.8]{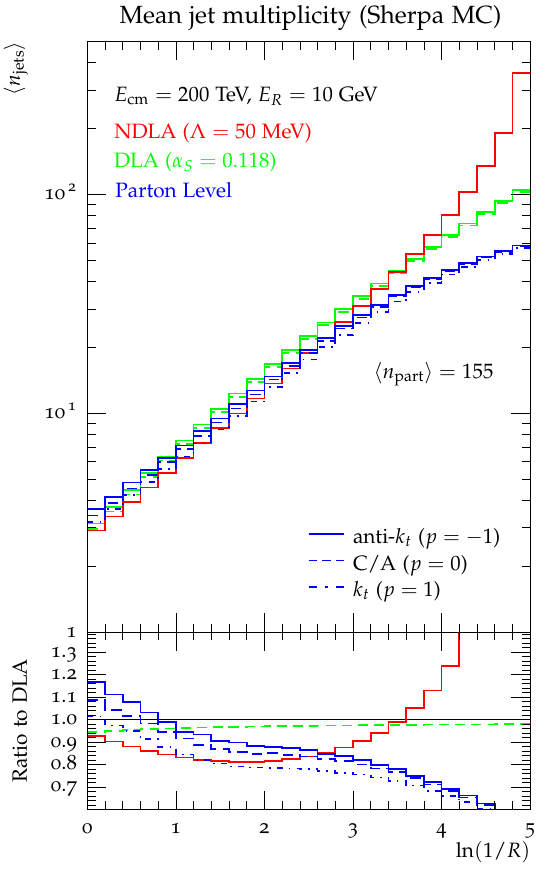}
  \caption{\label{fig:eesherpa}%
(Left) Average \Sherpa jet multiplicity in $\ee$ annihilation at 200 TeV,
compared to the DLA formula (\ref{eq:Avn_I0}) (green), and the NDLA (\ref{eq:Ngcd}) 
(red).  The (black) solid, dashed and dot-dashed curves 
show \Sherpa hadron level predictions in the anti-$k_t$, C/A 
and $k_t$ algorithms, respectively.  (Right) Same but compared 
with the parton level simulation (blue).}}
}
In Fig.~\ref{fig:eesherpa} results for the average jet multiplicity 
from the \Sherpa Monte Carlo are compared with the fixed-coupling 
DLA results (for fixed $\as = 0.118$) and the NDLA results for
running $\as$, the latter derived numerically as explained in
the previous Section.  The average jet multiplicity 
is shown for $\Ecm = 200$ TeV and $E_R=10$ GeV as a function of the 
jet radius parameter $R$. For these extreme values we expect the 
leading logarithms to dominate.  At such a high energy, the
asymptotic approximation  (\ref{eq:Avn_asy}) (green dashed) agrees well
with the exact DLA formula  (\ref{eq:Avn_I0}) (solid green), even at
$R\sim 1$.  

A small QCD scale $\Lambda\sim 50$ MeV in the NDLA formulae
gives reasonable agreement with the Monte Carlo results,
in which $\Lambda_{\overline{MS}}=180$ MeV.  Such a scale change is a
next-to-NDLA effect and therefore within the uncertainties.
The Monte Carlo results follow the formulae up to
$\ln(1/R)\sim 4$, after which the parton-level Monte Carlo result
approaches the parton multiplicity (about 150 at the shower cutoff
$Q_0\sim 1$ GeV), while the NDLA diverges towards the Landau pole
of the running coupling at $\ln(1/R)\sim 5$.
Surprisingly, the hadron-level result follows the fixed-coupling
DLA perturbative prediction further, up to around 100 jets at
$\ln(1/R)\sim 5$, before breaking away to saturate at the hadron
multiplicity.  We note that the quantitative agreement between Monte Carlo
simulation, implementing the running coupling, and the analytic DLA 
result depends on the choice for the fixed coupling in the latter. For 
the extreme c.m.\ energy we consider, a somewhat smaller value for $\as$ 
can improve the agreement with the simulation for larger values of 
$\ln(1/R)$. At the same time for lower values of $\ln(1/R)$, where 
we do not expect the DLA to fully apply, the agreement would get worse. 

Figures~\ref{fig:ecmsherpa_PL} and \ref{fig:ecmsherpa_HL} show \Sherpa parton 
and hadron level results for the jet multiplicity as a function of c.m.\ 
energy, compared with the DLA and NDLA predictions.
Again, a QCD scale $\Lambda\sim 50$ MeV in the NDLA formulae
gives reasonable agreement with the Monte Carlo results.
At the smallest values of $R$, the Monte Carlo values are approaching
the NDLA predictions from below, indicating further subleading
effects. As in Fig.~\ref{fig:eesherpa}, the agreement at very small $R$ is
better at hadron level than at parton level.

\FIGURE{
  \centering\centerline{
  \includegraphics[scale=0.55]{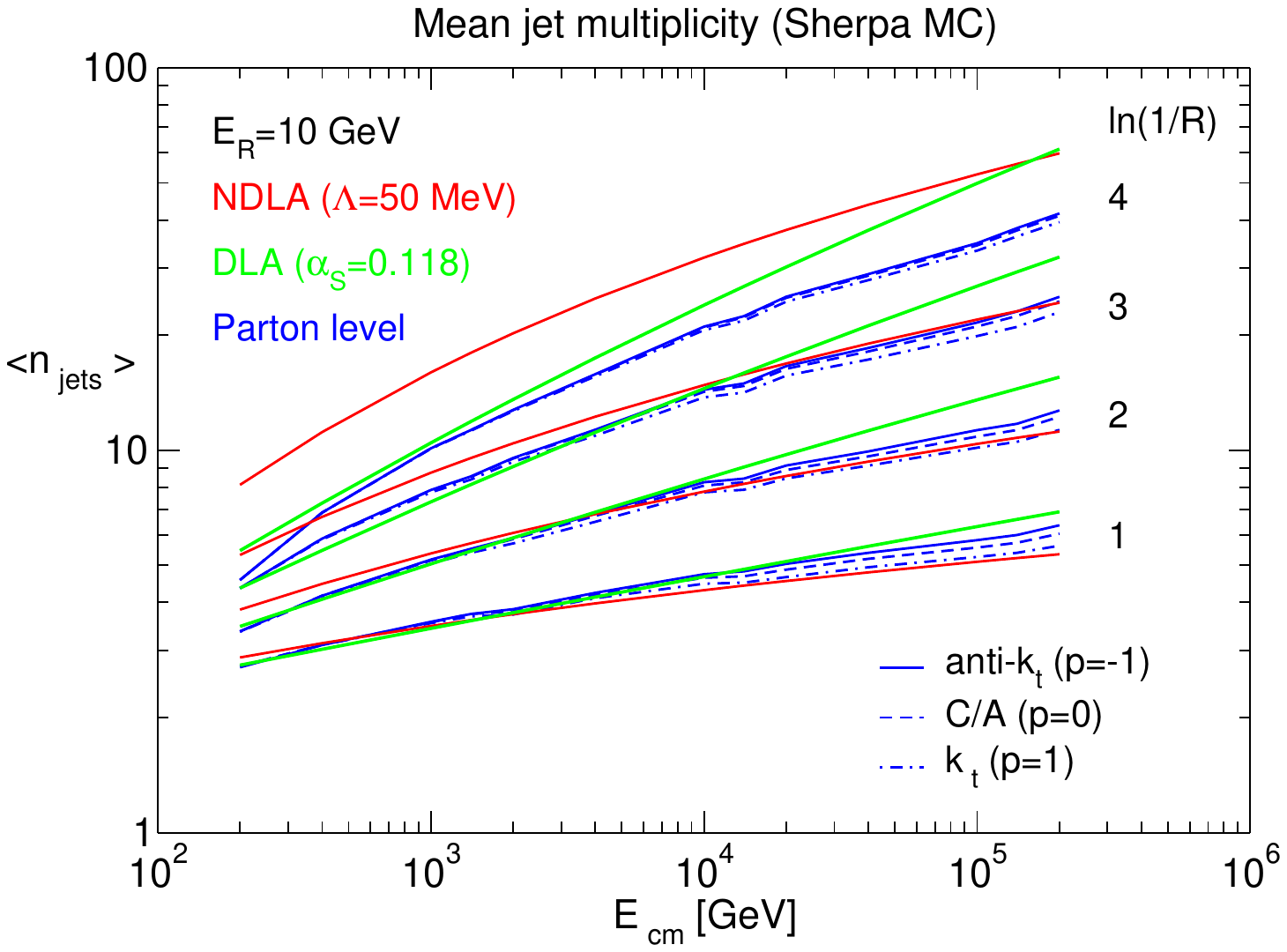}
  \caption{\label{fig:ecmsherpa_PL}%
Average \Sherpa jet multiplicity in $\ee$ annihilation as a
function of centre-of-mass energy, compared to the DLA (\ref{eq:Avn_I0}) (green) 
and NDLA (red) from Eqs.~(\ref{eq:dNqdkdl}) and (\ref{eq:dNgdkdl}).  
Solid, dashed and dot-dashed blue curves show \Sherpa parton 
level predictions for anti-$k_t$, C/A and $k_t$ algorithms, respectively.}}
}

\FIGURE{
  \centering\centerline{
  \includegraphics[scale=0.55]{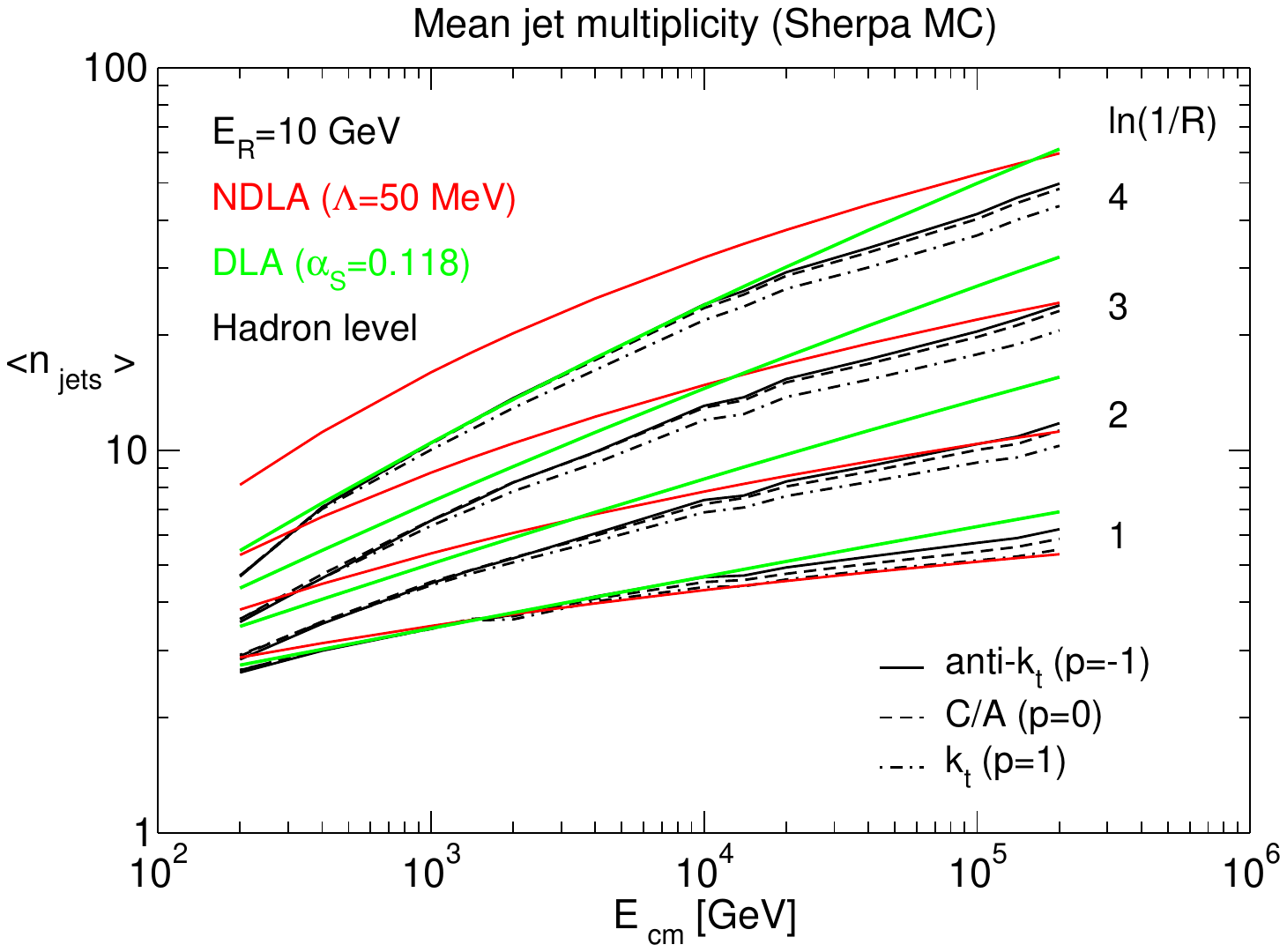}
  \caption{\label{fig:ecmsherpa_HL}%
Same as Fig. \ref{fig:ecmsherpa_PL} but compared with the hadron level (black).}}
}

It can be see from both figures that the Monte Carlo results for all three inclusive 
algorithms, anti-$k_t$, C/A and $k_t$, are very similar, consistent with 
the lack of dependence of our analytical predictions on the power $p$ in
Eqs.~(\ref{eq:genktee}) and (\ref{eq:genktpp}).  The jet multiplicity
is systematically slightly lower for the $k_t$ algorithm, which we
conjecture is due to its tendency to gather more low-momentum particles
into jets from angles somewhat larger than the canonical jet radius
$R$.  This leads to a smaller number of jets for a given final-state 
multiplicity. This higher susceptibility of $k_t$ jets to radiation was 
discussed in \cite{Cacciari:2008gn} where also a first-order estimate 
for the resulting effective area of $k_t$ and C/A jets was derived. 
Qualitatively we observe the same behaviour as for the passive jet 
area discussed there: the effect increases for higher jet energies 
and larger nominal $R$.

\section{\boldmath Monte Carlo results: $pp$\label{sec:mcpp}}
The motivation for studying the inclusive family of jet algorithms 
is the similarity to phenomenologically relevant hadron collider 
algorithms.  For jet rates, and therefore also the mean number 
of jets, the factorization of the PDF in the initial state holds at 
the double leading logarithmic order.  In this section we compare 
the predictions of our generating functions with Monte Carlo 
calculations and determine the reliability of our resummation in 
the LHC context.

\FIGURE{
  \centering\centerline{
  \includegraphics[scale=0.80]{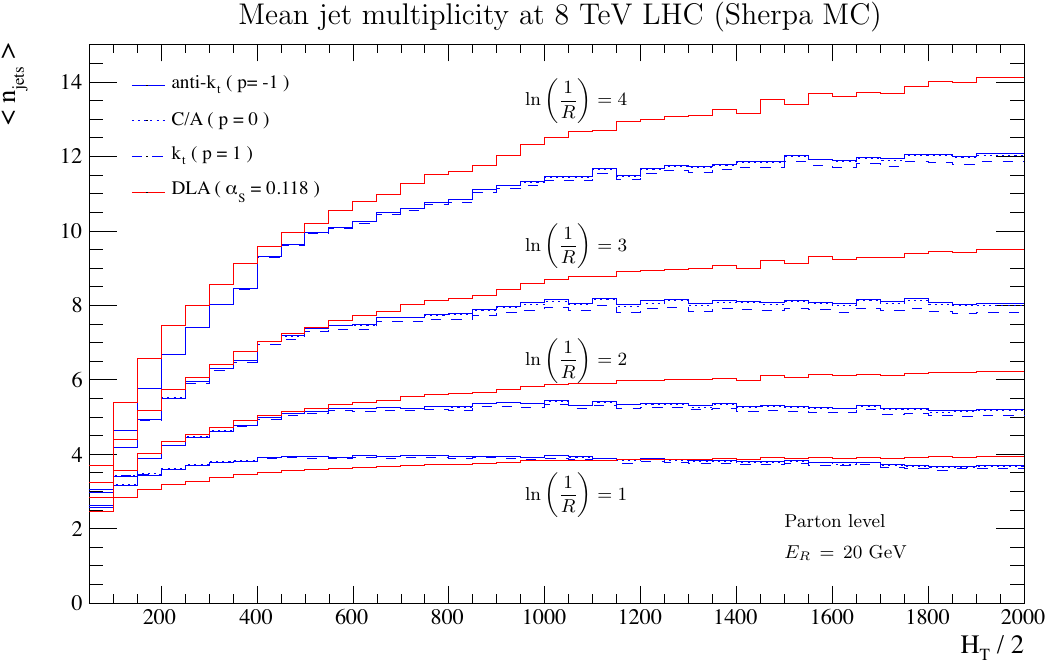}
  \vspace*{-20mm}
  \caption{\label{fig:hadronsS}%
    The proton-proton average jet multiplicity in inclusive di-jet production  
    as a function of $H_T/2$. Compared is the DLA analytic prediction, 
    shown in red, with parton shower results obtained with \Sherpa. The solid, 
    dashed and dot-dashed curves show anti-$k_t$, C/A and $k_t$ algorithms, 
    respectively.}}
}

Figure~\ref{fig:hadronsS} shows results for $pp$ collisions at the
current LHC energy of 8 TeV.  Here the average multiplicity of jets with 
$p_T>E_R=20$ GeV is plotted as a function of $H_T/2$, where 
\beq
H_T = \sum_i p_{T,i}\;,
\eeq
the scalar sum of all jet transverse momenta. This choice corresponds 
closely with the initial parton transverse momenta in the QCD $2\to 2$ 
hard scattering.  Each point in the analytic result is weighted by the 
partonic fraction of final-state quarks versus gluons.  In other words, 
the analytic result is
\beq
\VEV{n_{jj}} \; = \;2  \left( c_q\VEV{n_{q}} + c_g\VEV{n_{g}} \right),
\eeq
where $c_q + c_g = 1 $ and are determined from the proportions of final-state 
jets in the contributing $2\to 2$ hard subprocesses.  The considered values for 
$R$ are the same as in Fig.~\ref{fig:ecmsherpa_PL}.  The multiplicity levels 
off and even decreases at high values of the primary parton $p_T$, owing to 
the transition from gluon to quark jets. Overall we observe a good agreement
of the analytic estimate with the \Sherpa simulation. As before we use a 
fixed coupling of $\as=0.118$ in the DLA calculation. In particular for 
large values of $H_T/2 $ a reduced value could be more appropriate, improving 
the agreement with simulation. Notably, the dependence on the jet algorithm
for the simulated results is rather mild. As observed for $e^+e^-$ collisions
in the above, the $k_t$ algorithm tends to produce slightly less jets for a 
given final-state multiplicity. 

\FIGURE{
  \centering{
  \includegraphics[scale=0.70]{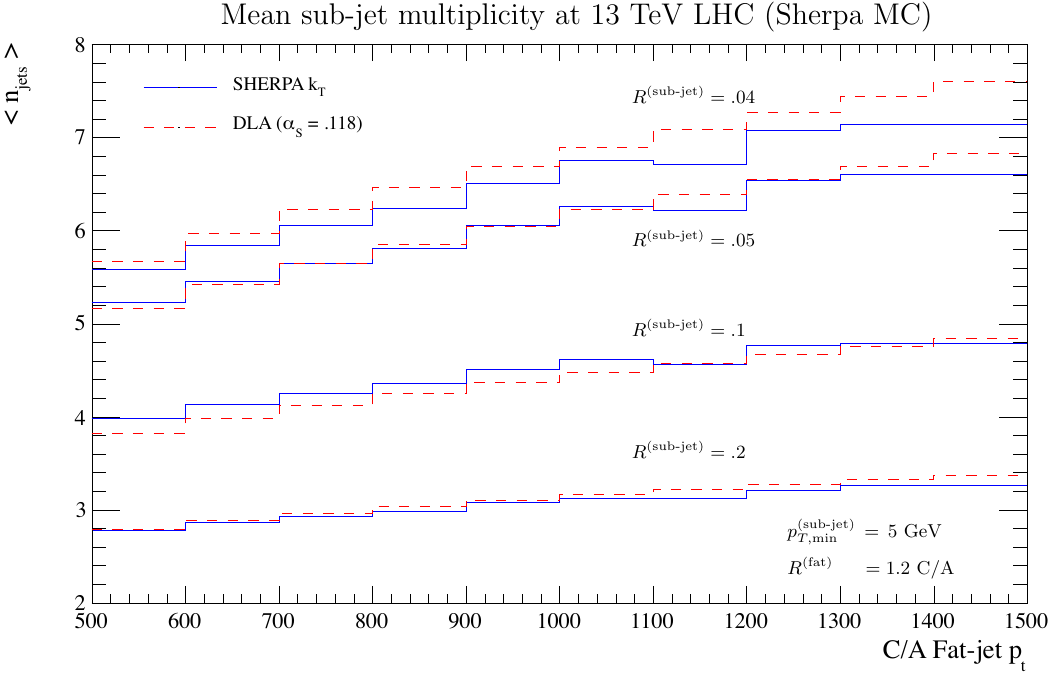}
  \caption{\label{fig:sub_jet_fat}%
Average sub-jet multiplicity for parton level \Sherpa $pp \to q\bar{q}$ sample 
compared with the modified DLA result.  Sub-jets,  defined as $k_{t}$-jets with 
radius $R^{(\text{sub-jet})} = (.2,.1,.05,.04)$, are counted inside a C/A 
fat-jet with radius $R^{(\text{fat})}=1.2$.
}
}}

\FIGURE{
  \centering{
  \includegraphics[scale=0.90]{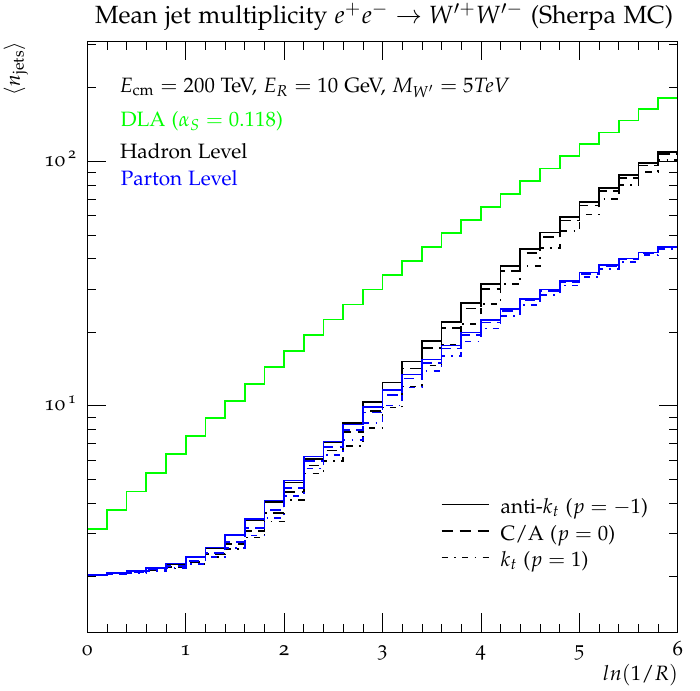}
  \caption{\label{fig:boost}%
Average jet multiplicity from \Sherpa for a boosted configuration 
of decay jets in $e^+ e^- \to W'W'$ compared with the DLA result 
assuming a pure QCD sample (green).  From the simulation we 
include the parton (blue) and hadron (black) results for the 
anti-$k_t$ (solid), C/A (dashed) and $k_t$ (dot-dashed) algorithms.
 }}
}

\section{Sub-jet multiplicities}

\label{sec:sjet}

Jet substructure is an increasing relevant handle for distinguishing 
strongly decaying new physics from the QCD background \cite{Altheimer:2012mn}.  
A particular observable of interest is the sub-jet ($R^{(\text{sub-jet})}\sim 0.1-0.4$) multiplicity 
inside a larger fat jet (typically with 
a $R^{(\text{fat})} \sim 1 - 1.2$)~\cite{Cohen:2012yc,Hedri:2013pvl}.  
In this section we briefly outline how our generating function can predict 
QCD sub-jet multiplicities, and how these vary significantly from new 
physics signals.

In the first case, we modify our formula for the average jet multiplicity 
to account for the reduced available phase space of the fat-jet.  
At the DLA, the phase space boundary in the $\xi$ integration is now the 
fat-jet radius $R^{(\text{fat})}$ as opposed to the entire hemisphere, 
so that the angular logarithm in our average jet multiplicity (\ref{eq:Avn_I0}) 
becomes $\log(R^{(\text{fat})}/R^{(\text{sub-jet})})$. Further, we choose the fat-jet 
$p_T$ as the upper energy scale in analytic formula. With these modifications 
we compare the analytic formula with a SHERPA sample of pure quark jets in 
Fig.~\ref{fig:sub_jet_fat}. The DLA result captures the scaling with respect 
to fat-jet $p_T$, $R^{(\text{fat})}$ and $R^{(\text{sub-jet})}$ quite well. Even so we have to
note, that the level of agreement is highly dependent on the chosen value for $\as$.   

As a second example we contrast in Fig.~\ref{fig:boost} our DLA multiplicity 
estimate with the result obtained for the production of a pair of hadronically 
decaying $W'$ bosons of mass $5~$TeV. We again consider a centre-of-mass energy 
of $200~$TeV resulting in rather boosted, collimated decays for the bosons. 
As a consequence we observe that in particular for sizable values of $R$, where the 
substructure of the $W'$ decays is not yet resolved, we expect many fewer jets than
predicted for pure QCD production. As we decrease $R$ we resolve more and more jets
again, the hadron level jet-multiplicity being significantly higher than the one at
parton level.

\section{Jet scaling patterns\label{sec:scale}}

\FIGURE{
  \centering\centerline{
  \includegraphics[scale=0.50]{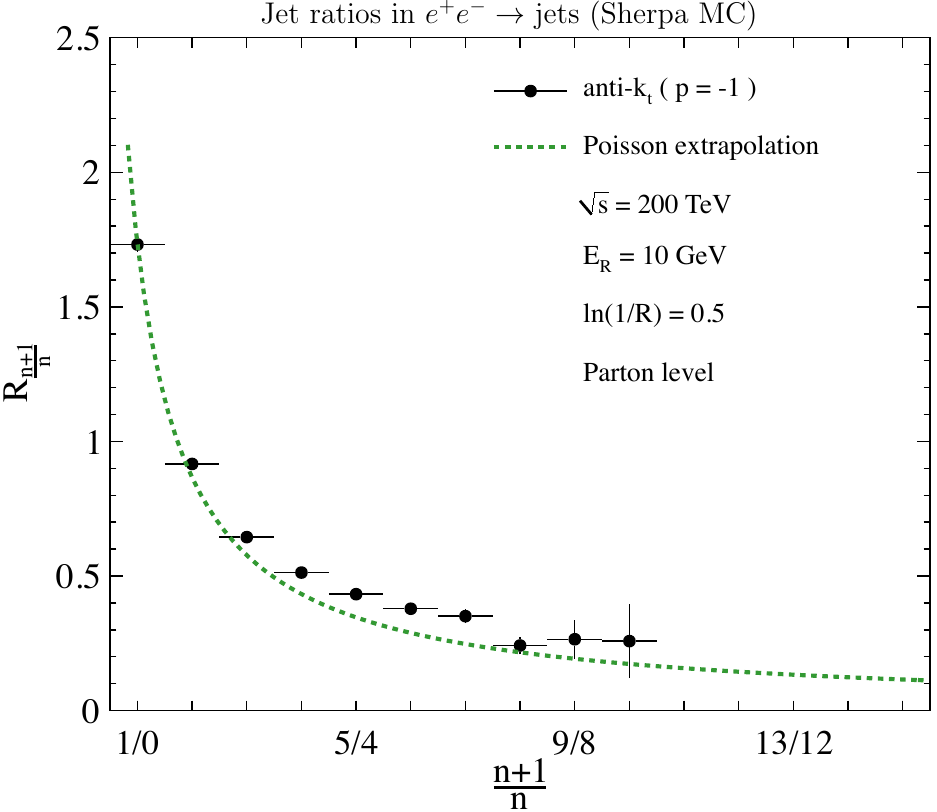}
  \includegraphics[scale=0.50]{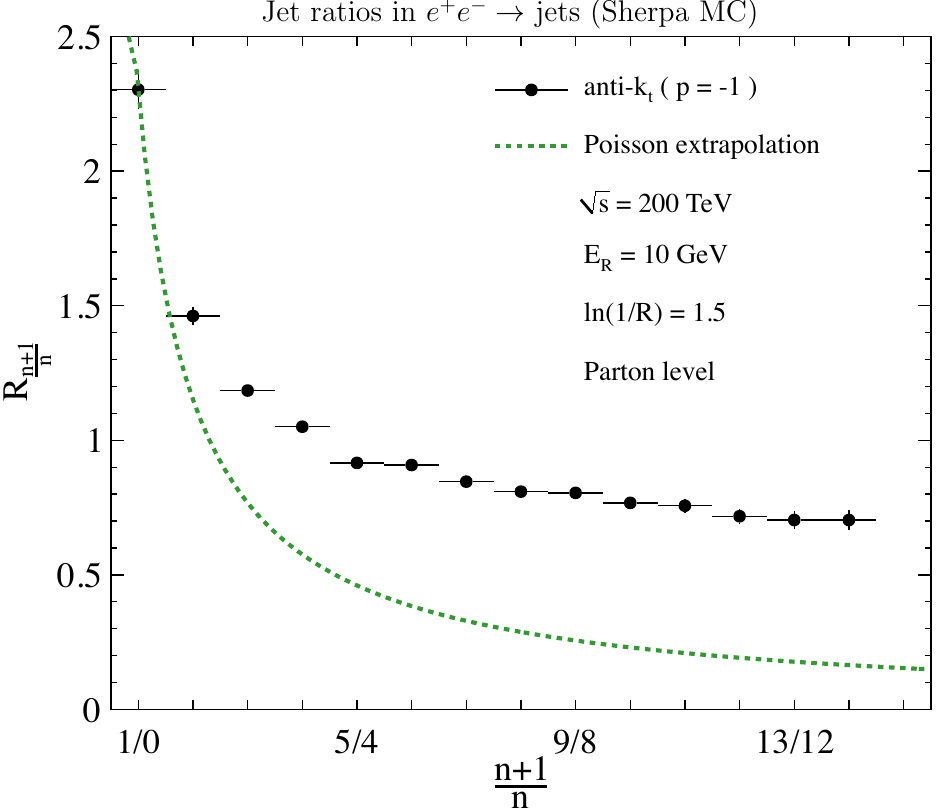}
  \caption{\label{fig:genratios}%
Jet ratios from \Sherpa compared with the Poisson extrapolation 
from the first bin.  For smaller jet sizes the Poisson behaviour 
breaks down as discussed in the text.  Jet sizes here correspond 
to approximate $R$
values of 0.61 (left) and 0.22 (right).}}
}

The work in \cite{Gerwick:2012hq,Gerwick:2011tm,Englert:2011pq} discussed the potential 
for extrapolating jet rates based on universal scaling 
patterns.  These patterns are most easily classified in 
terms of the ratio of exclusive jet rates $R_{(n+1)/n} = \sigma_{n+1}/\sigma_n$.  
In the Durham (exclusive $k_t$) algorithm, it was found that for low 
multiplicity $n \le \VEV{n_{jets}}$, emissions are essentially 
Poisson-like so that $R_{(n+1)/n} \sim (n+1)^{-1}$.  
The tail of the multiplicity distribution 
then produces dominantly staircase or geometric scaling 
where $R_{(n+1)/n} \sim \text{constant}$.  This regime 
is driven by the fractal nature of QCD radiation in the 
gluon dominated limit. 

We know from previous work that the expected scaling 
patterns of jets can depend dramatically on the jet algorithm.  One  
example of this is the JADE algorithm, where the non-exponentiation 
of the primary emissions precludes the Poisson extrapolation even 
in the pseudo-abelian limit \cite{Brown:1990nm}.
In this section we would like to address scaling in the inclusive generalized $k_t$ 
class of algorithms.  This extends the results in \cite{Gerwick:2012hq} 
and strengthens the case for investigation at hadron colliders.

\subsection{Poisson breaking components} With the leading 
logarithmic coefficients from  Eqs. (\ref{eq:R2})-(\ref{eq:R5}) it is 
easy to make some first statements about scaling in the generalized algorithm.  
It is clear from the structure of the coefficients that in the limit 
$C_A \to 0$ a perfect Poisson distribution emerges.  Now a 
simple comparison between the generalized
and Durham algorithms is the relative 
size of the Poisson breaking components in the lower 
multiplicity rates, for example the $2$-gluon 
correlated emission contribution to the $4$-jet rate (\ref{eq:R4}). 
For the double-leading logarithmic coefficients to 
the $4$-jet rate, $R_{44}$, we find
$C_{44}^{\text{Durham}} \sim 2C_F^2 + (1/3) C_A C_F$ and
$C_{44}^{\text{generalized}} \sim  2C_F^2 + (1/2) C_A C_F$
using the normalization of this work.  The lowest-order Poisson 
breaking term is relatively 
larger in the generalized class of algorithms.  We would thus 
expect the onset of staircase (geometric) scaling to come 
about for even smaller values of the logarithm, and to better 
match the staircase behaviour at lower multiplicity.  Evidence 
of this may be found in the $Z$ + jets analysis presented 
in \cite{Gerwick:2012hq}.  We present in table \ref{poscas} 
the relative sizes of the Poisson breaking terms in the 
DLA expanding coefficients for the two algorithms compared with 
the idealized Poisson and staircase predictions.  In this 
table we compute the 6-jet rate in the generalized algorithm by 
expanding the resummed jet rates from \eq{resummed_func5} 
to $\mathcal{O}(\alpha_s^4)$ which entirely determines 
the DLA resolved component.  \bigskip

\begin{table}[h]
\begin{center}
\begin{tabular}{|c|cccc|}
\hline
& Poisson & Generalized $k_t$ & Durham & Staircase \\
\hline 
$R_{4/3}/R_{3/2}$ & $0.50$  & $0.781$  & $0.688$  & $1$   
       \\[2mm] 
$R_{5/4}/R_{4/3}$ & $0.67$  & $0.906$  & $0.868$  & $1$ \\[2mm]
$R_{6/5}/R_{5/4}$ & $0.75$  & $0.923$  & $0.932$  & $1$ \\[2mm] 
\hline 
\end{tabular}
\end{center}
\caption{Ratio of successive ratios for the generalized $k_t$ and  
Durham jet rates compared with the idealized Poisson and staircase 
expectations.   } 
\label{poscas}
\end{table}
\bigskip
 
\subsection{$R$-dependence of scaling} 
A second question we wish to answer in this 
section is how the idealized scaling patterns 
depend on the jet radius parameter R in the 
generalized $k_t$ algorithm.  The additional handle 
provided by the separation of the angular and energy 
regulation allows us to probe an effect unbeknownst 
in the Durham algorithm.  At the double leading logarithmic 
level, the resummed rates in the general algorithm 
are invariant under exchange of $\k \leftrightarrow \l$.  
Decreasing the size of the jet merely increases the 
overall logarithm. 
  
In the simulation however, although we increase the overall 
logarithm when we require smaller resolved jets, we also 
change the relative contributions of the primary and 
secondary contributions due to kinematics.  This effect is present 
even at the level of the ordered $2$-gluon emission matrix 
element \cite{mat_ele}.

Using the parton shower we find that larger jet sizes 
dramatically increases the goodness of the fit with 
respect to the Poisson hypothesis.  This has a clear 
interpretation.  Correlated emissions in these events are 
predominantly intra-jet evolution for large jet radius.  In 
addition, this effect is larger than that due to the breaking 
of $\k \leftrightarrow \l$ symmetry at NDLA.

To analyze this further we investigate the observable $\theta_{34}$, defined as
\beq
\cos \theta_{34} \equiv { \bf \frac{p_3 \cdot p_4}{|p_3| |p_4|}},
\eeq	
in reconstructed $4$-jet events.  The third and fourth 
hardest jets overwhelmingly originate from 
emitted gluons.  As we suspect that in the simulation 
secondary Poisson breaking emissions are on average 
closer in $\theta_{34}$, this observable gives another 
estimate on the size of these breaking effects.  Comparing  
the prediction of pseudo-abelian QCD (in practice the parton 
shower with the $g\to gg$ and $g\to q\bar{q}$ branching 
off) we see in Fig.~\ref{fig:theta34} that the jet size has a 
large effect on the relative size of the correlated emission 
component.  In fact it appears that roughly $R \sim 0.4-0.5$ is 
the smallest value of the radius where the primary emissions 
are still the dominant contribution to the 4-jet rate. 

\FIGURE{
  \centering{
  \includegraphics[scale=0.78]{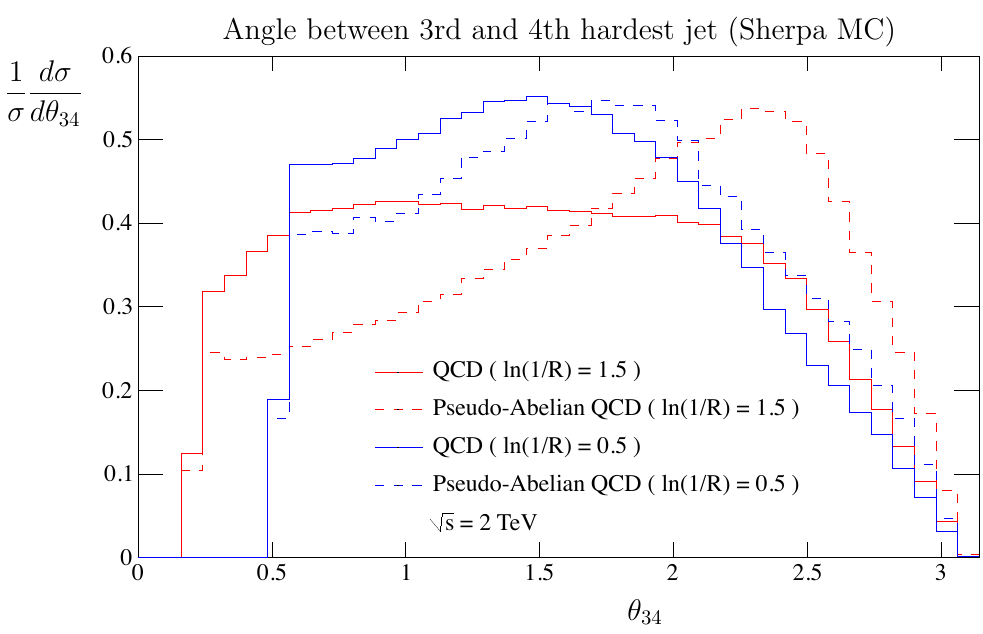}
  \caption{\label{fig:theta34}%
Differential cross-section with respect to the angle $\theta_{34}$ in 
exclusive 4-jet events.  The solid curves in both cases represents 
the parton shower while the dashed curves have the gluon splitting 
function turned off (both to quarks and gluons).    The large 
contribution of subsequent splittings between the red curves  
(small jet sizes) prevents a valid Poisson extrapolation as 
seen in Fig. \ref{fig:genratios}.    }}
}

\section{Conclusions}\label{sec:conc}
In the present paper we have derived the generating functions of jet
rates for the inclusive generalized $k_t$ jet algorithms, valid in the
next-to-double log approximation (NDLA), and used them to compute
jet rates and average jet multiplicities as functions of the jet radius
parameter $R$ and the minimum jet energy $E_R$.  At this level of
precision, the results are independent of the power $p$ that
distinguishes the inclusive $k_ t$, Cambridge/Aachen and anti-$k_ t$
algorithms.

The analytical results on $\ee$ annihilation are in broad agreement
with those from the \Sherpa Monte Carlo event generator, including the
weak $p$-dependence. Surprisingly, the hadron-level Monte Carlo
results follow the analytical predictions down to very small jet
radii, well beyond the range of the perturbative parton shower,
indicating that the cluster hadronization model in \Sherpa is smoothly
matched to perturbation theory.  Analytical predictions for $pp$
collisions at the current LHC energy also agree fairly well with
\Sherpa Monte Carlo results.

The \Sherpa Monte Carlo results and generating function in the generalized 
$k_t$ algorithm were also used to study the transition from Poisson-like to 
staircase-like behaviour in the $(n+1)/n$ jet ratios.  The relatively larger 
non-abelian terms in the DLA $\ee$ jet rates, compared to the Durham algorithm, 
indicate a transition at lower values of $n$.  Furthermore, the jet radius was 
found to have a significant impact on scaling, where smaller jet sizes 
receive larger secondary contributions.  This feature may be relevant for 
jet activity studies in jet substructure analyses.

These results represent the most detailed comparison to date between
analytical and Monte Carlo predictions for the inclusive jet
algorithms in current use at the LHC.  Information about the
dependence of jet rates and multiplicities at high jet energies and
relatively small radii are particularly important for studies of
boosted jets and jet substructure, which play an increasingly
important role in searches for new physics at the LHC.  As we 
demonstrated, our leading-logarithmic predictions are relevant 
to such studies, and their fairly good agreement with Monte Carlo 
results is encouraging.  Possible improvements such as optimized 
scale choices and empirical subleading contributions would 
be worth exploring.  Studies of other features of the jet multiplicity 
distribution along these lines could also be useful.

\section*{Acknowledgements}
BW acknowledges the support of a Leverhulme Trust Emeritus Fellowship,
and thanks the Pauli Institute at ETH/University of Zurich
for hospitality and support during part of this work. EG and SS acknowledge
support by the Bundesministerium f\"ur Bildung und Forschung under contract 
05H2012.

\appendix
\section{Derivation of the PDE for the average jet multiplicity}
\label{sec:PDEderiv}
Differentiating \eq{eq:Nq} with respect to $\xi$ and using (\ref{eq:askt}), we
have to NDLA
\beqn
\xi\frac{\pd}{\pd\xi}\frac{\N_q-1}{\bas} &=& C_F\int_{E_R/E}^1 dz
\left(\frac 1z -\frac 34 -b_0\bas\frac 2z\ln z\right)\N_g(zE,\xi)\nn
&& + C_Fb_0\bas\int_{\xR}^{\xi} \frac{d\xi'}{\xi'}
\int_{E_R/E}^1 \frac{dz}z \N_g(zE,\xi')\;.
\eeqn
But at the same level of precision we have (remembering the
$\xi$ dependence of $\bas$)
\beq
\xi\frac{\pd}{\pd\xi}\frac{\N_q-1}{\bas} =\frac
1{\bas}\xi\frac{\pd\N_q}{\pd\xi} +b_0(\N_q-1)
\eeq
and
\beq
 C_F\bas\int_{\xR}^{\xi} \frac{d\xi'}{\xi'}
\int_{E_R/E}^1 \frac{dz}z \N_g(zE,\xi')=\N_q-1\;,
\eeq
so that
\beq
\xi\frac{\pd\N_q}{\pd\xi} = C_F\bas\int_{E_R/E}^1 dz
\left(\frac 1z -\frac 34 -b_0\bas\frac 2z\ln z\right)\N_g(zE,\xi)\;.
\eeq
Differentiating with respect to $E$ and using (\ref{eq:askt}) again,
\beqn
E\frac{\pd}{\pd E}\frac{\xi}{\bas}\frac{\pd\N_q}{\pd\xi} &=& C_F\N_g
+\frac 34 C_F\int_{E_R/E}^1 dz\left[\N_g(zE,\xi)-\N_g(E,\xi)\right]\nn
&&+2C_Fb_0\bas\int_{E_R/E}^1 \frac{dz}z\N_g(zE,\xi)\;.
\eeqn
Now to the required precision
\beq
\int_{E_R/E}^1 dz\left[\N_g(zE,\xi)-\N_g(E,\xi)\right]
=E\frac{\pd\N_g}{\pd E}\int_0^1 dz\ln z =-E\frac{\pd\N_g}{\pd E}
\eeq
and
\beq
2C_Fb_0\bas\int_{E_R/E}^1 \frac{dz}z\N_g(zE,\xi)
=2b_0\xi\frac{\pd\N_q}{\pd\xi}
\eeq
while (remembering the $E$ dependence of $\bas$)
\beq
E\frac{\pd}{\pd E}\frac{\xi}{\bas}\frac{\pd\N_q}{\pd\xi} 
=\frac 1{\bas}E\frac{\pd}{\pd E}\xi\frac{\pd\N_q}{\pd\xi} 
+2b_0\xi\frac{\pd\N_q}{\pd\xi}\;.
\eeq
Thus we get the PDE
\beq
E\frac{\pd}{\pd E}\xi\frac{\pd\N_q}{\pd\xi} 
=C_F\bas\left(\N_g-\frac 34 E\frac{\pd\N_g}{\pd E}\right)\;,
\eeq
which gives (\ref{eq:dNqdkdl}) in terms of the logarithmic variables (\ref{eq:kldef}).

\section{Properties of the PDE for the average jet multiplicity}
\label{sec:PDEprops}
The PDE (\ref{eq:NgPDE}) is not so straightforward to solve explicitly. If we consider changing
to variables corresponding to the sum and difference of $\k$ and $\l$,
we find that the PDE is separable (one obtains ODEs that are
equivalent to those describing the classical mechanics of the harmonic
oscillator and the quantum mechanics of the Coulomb potential), but
that the boundary conditions are not. As a result, one may prove that
one cannot express the boundary conditions as a Fourier-Bessel series
of orthogonal functions of Sturm-Liouville type.

On reflection, seeking an explicit solution in this way is perhaps ambitious,
given that there is no guarantee that solutions to such a PDE can be
written in terms of special functions. 
We thus proceed to an analysis of a rather different nature. This
analysis will enable us to (i) establish that a solution
exists; (ii) give a variety of infinite montonic series of upper and lower
bounds
on that solution; and (iii) provide explicit series solutions that
enable, {\em e.g.}, the asymptotic behaviour of the solution to be
found in simple, closed form. 

To proceed, it is useful to define 
$ x=\frac{c_g}{2} (2\k+\frac{\mu}{2})$, $
y=\frac{c_g}{2}(\l+\frac{\mu}{2})$, in terms of which
\beq \label{eq:npde}
\N_{xy} = \frac{\N}{x+y}.
\eeq
 We may recast
the PDE in the integral forms
\beq
\frac{\pd \N}{\pd y} = \int_a^x dx^\prime \frac{\N}{x^\prime + y^\prime}\,,
\eeq
where $a= \frac{c_g \mu}{4}$, or 
\beq
\N = 1+ \int_a^y dy^\prime \int_a^x dx^\prime \frac{\N}{x^\prime + y^\prime}\,.
\eeq
The first of these forms, by the way, makes clear that a
physically-acceptable solution, $\N$, if it exists, is a monotonically
increasing function of $x, \forall \; y,$ and of $y, \forall \; x$. (Proof: The mean number of jets
must be $\geq 0$. Thus $\N_y >0$; $\N_x
>0$ follows by symmetry in $x \leftrightarrow y$.) 

The basic idea behind our analysis will be to first identify related
PDEs for which we can find explicit solutions whose properties (such
as their existence, continuity, {\em \&c.}) can be
checked `by hand'. We then use these solutions as a crutch to derive
properties (for example, the existence) of the solution of the
original PDE.

Our first lemma is as follows. Suppose $\M^0 (x,y)$ is a solution of
the PDE
\beq \label{eq:mpde}
\M^0_{xy} = \alpha (x,y) \M^0
\eeq
subject to the same boundary conditions as $\N$. Suppose furthermore
that $\alpha (x,y) \geq \frac{1}{x+y}$ almost
everywhere\footnote{We use the usual language of measure theory.} in the
domain $D \equiv \{(x,y)|x,y \geq a\}$. Then
\beq
\M^1 (x,y) \equiv 1 + \int_a^y dy^\prime \int_a^x dx^\prime
\frac{\M^0}{x^\prime + y^\prime} \leq \M^0, \; \forall \; (x,y) \in D.
\eeq
( For brevity, we denote a double integral of the type that
appears on the RHS
by $\int \frac{\M^0}{x+y}$ henceforth.) Proof of this follows immediately from
the fact that $1+ \int
\frac{\M^0}{x+y} \leq 1+ \int
\alpha \M^0 =\M^0$ everywhere in $D$. Continuing in this vein, we
define an infinite sequence whose elements are 
\beq \label{eq:mseq}
\M^{i+1} (x,y) \equiv 1 + \int 
\frac{\M^i}{x+y} \,.
\eeq
Now
\beq
\M^{i+1}- \M^i = \int 
\frac{\M^i-\M^{i-1}}{x+y} \,
\eeq
so $\M^{i}- \M^{i-1} \leq
0 \; \forall (x,y) \in D \implies \M^{i+1}- \M^i \leq 0 \; \forall (x,y) \in D$ and proof that this sequence is montonically
decreasing follows by induction, since we have already proven that
$\M^1 \leq \M^0$ everywhere in $D$.

We next prove that the sequence $\{\M^i \}$ is bounded below by zero,
if $\M^0 > 0$ (which we can verify by hand given an explicit $\M^0$),
and hence converges. To do so, we simply note that positivity of
$\M^0$ implies positivity of $\M^i \; \forall \; i$ by induction, given
(\ref{eq:mseq}) and $\M^0 > 0$.

We now prove that the sequence $\{\M^i\}$ converges to a solution,
$\overline{\N}$ of the original PDE. Indeed
\beq
\overline{\N} \equiv \lim_{i \rightarrow \infty} \M^i= \lim_{i \rightarrow \infty} \M^{i+1} = 1 + \lim_{i \rightarrow
  \infty}
\int \frac{\M^i}{x+y} = 1 +
\int  \lim_{i \rightarrow
  \infty} \frac{\M^i}{x+y} = 1+ \int \frac{\overline{\N}}{x+y},
\eeq
where, in the penultimate step we used the Theorem of Monotone Convergence of Lebesgue integration.\footnote{This Theorem requires that
  the $\M^i$ be measurable, but this follows from the fact $M^0$, and
  hence $\M^i$ are $C^0$. Note, however, that we have not proven that
  $\overline{\N}$ itself is $C^0$, let alone $C^1$ or
  $C^2$. Strictly speaking, therefore, we have proven
  that $\overline{\N}$ is a solution of the integral equation, which
  is what we started with, rather
  than the PDE. We shall in any case ignore this subtlety in what follows.}

We shall call $\overline{\N}$ a {\em supersolution} of
(\ref{eq:npde}). Evidently, given any suitable $\alpha$ and
$\M^0$, we can use the above results to prove that a solution exists and
to find an infinite series of monotonically decreasing upper bounds on
it. This raises the question of uniqueness, however: different starting
points, $(\alpha,\M^0)$, may lead to different supersolutions.

It is straightforward to prove the following weak version of
uniqueness, for sequences that are `nested' in the following sense.
Suppose two sequences $\{\M^i\}$ and $\{\M^{\prime i}\}$ converge on
supersolutions $\overline{\N}$ and $\overline{\N}^\prime$,
respectively. If $\exists \; i,j,k,l$ such that elements $\M^i \geq
\M^{\prime j}$ and $\M^k \leq
\M^{\prime l}$  almost everywhere, then $\overline{\N}
=\overline{\N}^\prime$. Proof: $\M^i \geq
\M^{\prime j} \implies \M^i \geq
\overline{\N}^\prime$. But $\M^i -
\overline{\N}^\prime \geq 0 \implies \int \frac{\M^i - \overline{\N}^\prime}{x+y} =
\M^{i+1} - \overline{\N}^\prime \geq 0 \implies  \overline{\N} \geq
\overline{\N}^\prime$. Similarly, $\M^k \leq
\M^{\prime l} \implies \overline{\N} \leq
\overline{\N}^\prime \implies \overline{\N} =
\overline{\N}^\prime$, QED. Unfortunately, this weak version of
uniqueness does not even imply that any two supersolutions coincide,
since the sequences that lead to them may be wholly contained in
distinct intervals, or, even if they are nested, an element of one
sequence need not exceed an element of the other almost everywhere.
But this weak version is still useful where it applies, in that it may give us a
more rapidly converging sequence of upper bounds on a solution.

Let us now discuss lower bounds. 
Similar arguments to those just given
show that if we instead started with a solution, $\L^0 (x,y)$ to a PDE of the form
\beq \label{eq:lpde}
\L^0_{xy} = \beta (x,y) \L^0,
\eeq
with $\beta < \frac{1}{x+y}$ almost everywhere on $D$, then we shall obtain
an infinite, monotonically {\em increasing} sequence.
We can, furthermore, prove that this sequence is
  bounded above 
and hence converges, provided we can show by hand that
 $\L^0 \leq \M^0$, for some $\M^0$ that converges to a
 supersolution $\overline{\N}$. To wit, suppose we have shown that
 $\L^0 \leq \M^0$. It then follows by induction that $\L^i \leq \M^i$,
 since $\L^{i+1} - \M^{i+1} = \int \frac{\L^{i} - \M^{i}}{x+y}$, and
 thus that $\lim_{i \rightarrow \infty} \L^i \leq
 \overline{\N}$. Again, by the Theorem of Monotone Convergence, $\{
 \L^{i} \}$ converges to a solution of (\ref{eq:npde}) that we call a
 {\em subsolution} and denote by $\underline{\N}$. By a
 straightforward generalization of the result just
 proven, a subsolution is less than or equal to any supersolution for
 which it can be shown that any one element in the sequence defining the
 former is less than or equal to (almost everywhere) any element in the sequence defining
 the latter. 

Yet again, this result is insufficient to establish uniqueness, but it
does provide a way to obtain both upper and lower bounds on a given solution. 
Obviously, if we can prove (or assume) uniqueness, then all of the
aforementioned super- and sub-solutions coincide, meaning that any of
the aforementioned sequences can be used to bound the solution to
arbitrary, known precision.

We can prove uniqueness of the solution heuristically out to any finite
$(x,y)$ by explicit
construction. Given $\N =1$ on the boundary, we divide the intervals
$[a,x]$ and $[a,y]$ into regions of size $\delta x$ and $\delta y$ and
find the unique solution $\N (a + \delta x,y) \simeq 1 + \delta x \int_a^y
\frac{dy^\prime}{a+y^\prime}$, with a similar result for $\N (x,a +
\delta y)$. We then extrapolate to the region $[a+\delta x,a+2\delta
x]$ and so on. Given that we have proven the existence of the
solution, we do not need to worry that the extrapolations in the $x$
and $y$ directions might not coincide. To make this proof rigorous
outside a neighbourhood of the boundary, 
we would then have to take the limit $\delta x, \delta y \rightarrow
0$.  We will content ourselves with assuming that the solution is unique. 

As we stressed above, all of our results on super- and sub-solutions are contingent on showing the existence of
solutions to the equations (\ref{eq:mpde}) and (\ref{eq:lpde}), for
suitable $\alpha$ and $\beta$, satisfying the BCs, and showing that
they have the desired properties. We do this by supplying explicit
solutions for functions $\alpha, \beta$ of the combination
$(x-a)(y-a)$, for which the PDE reduces to an ODE, and for which the
boundary conditions reduce to a single boundary condition at $(x-a)(y-a)=0$.

To obtain lower bounds, once may start, for example, with the solution
$\L^0 = 1$ of the PDE $\L^0_{xy} = 0$, since $\frac{1}{x+y} >0$
everywhere in $D$. We thus obtain 
\beq
\L^1 \geq 1+  (x+y) \ln (x+y) -  (x+a) \ln (x+a) - (a+y) \ln (a+y) + 2a \ln 2a\,,
\eeq
with subsequent $\L^i$ given in terms of polylogarithms.
At fixed $y$ and large
$x$, for example, we deduce that $\L^1 \sim (y-a) \ln x$.
More generally, $\L^n \sim \frac{((y-a) \ln x)^n}{(n!)^2} \implies
\N \sim I_0 (y-a) \ln x) \sim \frac{x^{(y-a)}}{\sqrt{2\pi (y-a) \log x}}$.

To find an upper bound, we can use $\alpha = = \frac{1}{2a}$ which evidently exceeds $\frac{1}{x+y}$
throughout $D$. This yields an initial upper bound of the form
\beq 
\M^0 = I_0 (\sqrt{2a(x-a)(y-a)}).
\eeq
A more stringent upper bound may be obtained from $\alpha =
\frac{1}{2\sqrt{(x-a)(y-a)}} > \frac{1}{x+y}$,
with solution 
\beq 
\M^0 = I_0 (2\sqrt{2}(x-a) ^{\frac{1}{4}} (y-a)^{\frac{1}{4}}).
\eeq

For a better lower bound, we can, provided $a>1$,
solve
\beq 
\L^0_{xy} = \frac{2}{(x+y)^2} \L^0,
\eeq
whose solution is
\beq 
\L^0 = \frac{xy + a^2}{a(x+y)}.
\eeq
It is soothing to verify explicitly that the resulting subsolution has
the same asymptotic behaviour, at large $x$ and fixed $y$, as the subsolution starting from $L^0
=1$ derived above. Indeed, in this limit, $\L^0  \sim \frac{y}{a}$,
$\L^1 \sim \frac{y^2-a^2}{2a}\log x$, and so on, with $\L^n \sim
\frac{(y-a)^n(y+na)}{a(n+1)!}\frac{\log x}{n!}$.
At large $n$, we thus find $\L^n \sim
\frac{(y-a)^n}{n!}\frac{\log x}{n!}$, exactly as before.


\begin{thebibliography}{10}
\bibitem{Salam:2009jx}
  G.~P.~Salam,
  ``Towards Jetography,''
  Eur.\ Phys.\ J.\ C {\bf 67} (2010) 637
  [arXiv:0906.1833 [hep-ph]].

\bibitem{Catani:1991hj}
  S.~Catani, Y.~L.~Dokshitzer, M.~Olsson, G.~Turnock and B.~R.~Webber,
  ``New clustering algorithm for multi-jet cross sections in $\ee$ annihilation,''
  Phys.\ Lett.\ B {\bf 269} (1991) 432.

\bibitem{Catani:1993hr}
  S.~Catani, Y.~L.~Dokshitzer, M.~H.~Seymour and B.~R.~Webber,
  ``Longitudinally invariant $k_t$ clustering algorithms for hadron hadron collisions,''
  Nucl.\ Phys.\ B {\bf 406} (1993) 187.

\bibitem{Ellis:1993tq}
  S.~D.~Ellis and D.~E.~Soper,
  ``Successive combination jet algorithm for hadron collisions,''
  Phys.\ Rev.\ D {\bf 48} (1993) 3160
  [hep-ph/9305266].

\bibitem{Cacciari:2008gp}
  M.~Cacciari, G.~P.~Salam and G.~Soyez,
  ``The anti-$k_t$ jet clustering algorithm,''
  JHEP {\bf 0804} (2008) 063
  [arXiv:0802.1189 [hep-ph]].

\bibitem{Cacciari:2011ma}
  M.~Cacciari, G.~P.~Salam and G.~Soyez,
  ``FastJet user manual,''
  Eur.\ Phys.\ J.\ C {\bf 72} (2012) 1896
  [arXiv:1111.6097 [hep-ph]].

\bibitem{Konishi:1979cb}
  K.~Konishi, A.~Ukawa and G.~Veneziano,
  ``Jet Calculus: A Simple Algorithm for Resolving QCD Jets,''
  Nucl.\ Phys.\ B {\bf 157} (1979) 45.

\bibitem{Dokshitzer:1991wu}
  Y.~L.~Dokshitzer, V.~A.~Khoze, A.~H.~Mueller and S.~I.~Troian,
  ``Basics of perturbative QCD,''
  Gif-sur-Yvette, France: Ed. Frontieres (1991) 274 p. (Basics of)

\bibitem{Ellis:1991qj}
  R.~K.~Ellis, W.~J.~Stirling and B.~R.~Webber,
  ``QCD and collider physics,''
  Camb.\ Monogr.\ Part.\ Phys.\ Nucl.\ Phys.\ Cosmol.\  {\bf 8} (1996) 1.

\bibitem{Gleisberg:2008ta}
  T.~Gleisberg, S.~.Hoeche, F.~Krauss, M.~Schonherr, S.~Schumann, F.~Siegert and J.~Winter,
  ``Event generation with SHERPA 1.1,''
  JHEP {\bf 0902} (2009) 007
  [arXiv:0811.4622 [hep-ph]].

\bibitem{Schumann:2007mg}
  S.~Schumann and F.~Krauss,
  ``A Parton shower algorithm based on Catani-Seymour dipole factorisation,''
  JHEP {\bf 0803} (2008) 038
  [arXiv:0709.1027 [hep-ph]].

\bibitem{CamOrig}
  Y.~L.~Dokshitzer, G.~D.~Leder, S.~Moretti and B.~R.~Webber,
  ``Better jet clustering algorithms,''
  JHEP {\bf 9708}, 001 (1997)
  [hep-ph/9707323];

\bibitem{CamWobisch}
  M.~Wobisch and T.~Wengler,
   ``Hadronization corrections to jet cross sections in deep-inelastic
  scattering,''
  arXiv:hep-ph/9907280;
  M.~Wobisch,
   ``Measurement and QCD analysis of jet cross sections in deep-inelastic
  positron proton collisions at s**(1/2) = 300-GeV,''
DESY-THESIS-2000-049.

\bibitem{Webber:2010vz}
  B.~R.~Webber,
  ``QCD jets and parton showers,''
  arXiv:1009.5871 [hep-ph].

\bibitem{Catani:1991pm}
  S.~Catani, Y.~L.~Dokshitzer, F.~Fiorani and B.~R.~Webber,
  ``Average number of jets in $e^+ e^-$ annihilation,''
  Nucl.\ Phys.\ B {\bf 377} (1992) 445.

\bibitem{Cacciari:2008gn}
  M.~Cacciari, G.~P.~Salam and G.~Soyez,
  ``The Catchment Area of Jets,''
  JHEP {\bf 0804} (2008) 005
  [arXiv:0802.1188 [hep-ph]].

\bibitem{Altheimer:2012mn} 
  A.~Altheimer, S.~Arora, L.~Asquith, G.~Brooijmans, J.~Butterworth, M.~Campanelli, B.~Chapleau  and A.~E.~Cholakian {\it et al.},
  ``Jet Substructure at the Tevatron and LHC: New results, new tools, new benchmarks,''
  J.\ Phys.\ G {\bf 39}, 063001 (2012)
  [arXiv:1201.0008 [hep-ph]].
  
\bibitem{Cohen:2012yc} 
  T.~Cohen, E.~Izaguirre, M.~Lisanti and H.~K.~Lou,
  ``Jet Substructure by Accident,''
  arXiv:1212.1456 [hep-ph].
  
\bibitem{Hedri:2013pvl} 
  S.~E.~Hedri, A.~Hook, M.~Jankowiak and J.~G.~Wacker,
  ``Learning How to Count: A High Multiplicity Search for the LHC,''
  arXiv:1302.1870 [hep-ph].
  
\bibitem{Gerwick:2012hq} 
  E.~Gerwick, T.~Plehn, S.~Schumann and P.~Schichtel,
  ``Scaling Patterns for QCD Jets,''
  JHEP {\bf 1210}, 162 (2012)
  [arXiv:1208.3676 [hep-ph]].
  
\bibitem{Gerwick:2011tm} 
  E.~Gerwick, T.~Plehn and S.~Schumann,
  ``Understanding Jet Scaling and Jet Vetos in Higgs Searches,''
  Phys.\ Rev.\ Lett.\  {\bf 108}, 032003 (2012)
  [arXiv:1108.3335 [hep-ph]].
  
\bibitem{Englert:2011pq}
  C.~Englert, T.~Plehn, P.~Schichtel and S.~Schumann,
  ``Establishing Jet Scaling Patterns with a Photon,''
  JHEP {\bf 1202} (2012) 030
  [arXiv:1108.5473 [hep-ph]].

\bibitem{Brown:1990nm} 
  N.~Brown and W.~J.~Stirling,
  ``Jet cross-sections at leading double logarithm in e+ e- annihilation,''
  Phys.\ Lett.\ B {\bf 252}, 657 (1990).
  
  \bibitem{mat_ele} 
  A.~Banfi, G.~Corcella and M.~Dasgupta,
  ``Angular ordering and parton showers for non-global QCD observables,''
  JHEP {\bf 0703}, 050 (2007).

\end{thebibliography}
\end{document}